\documentclass[twocolumn,tighten,twocolappendix]{aastex631}
\usepackage{times}
\usepackage{tabularx,booktabs}
\usepackage{mathtools}
\usepackage{enumitem}
\usepackage{gensymb}
\usepackage{xcolor}
\definecolor{lightgray}{rgb}{0.85,0.85,0.85}  
\usepackage{colortbl}
\usepackage{times}
\usepackage{xcolor}

\def\hi{{{\rm H}\,{\sc i}~}}
\def\hin{{{\rm H}\,{\sc i}}}
\def\cmsq{{{\rm cm$^{-2}$}}}
\def\hin{{{\rm H}\,{\sc i}}}

\def\msun{{M$_{\odot}$}}
\def\deg{$^{\circ} $ }
\def\teff{{\rm t$_{\rm eff}$}~}

\def\Tsys{{\rm T$_{\rm sys }$}~}

\defcitealias{Pingel2018}{P18}
\defcitealias{Das2020b}{D20}
\defcitealias{Das2024a}{D24}

\begin{document}

\title{Diffuse \hi emission in the circumgalactic medium of NGC\,891 and NGC\,4565 - III: azimuthal profiles}

\correspondingauthor{Mary Rickel}
\email{mrickel@nd.edu}
\author[0009-0007-9943-1183]{Mary Rickel}
\affiliation{Department of Physics \& Astronomy, University of Notre Dame, Notre Dame, IN USA}
\affiliation{Department of Astronomy, The Ohio State University, 140 West 18th Avenue, Columbus, OH 43210, USA}
\author[0000-0002-9069-7061]{Sanskriti Das}
\altaffiliation{Hubble Fellow}
\affil{Kavli Institute for Particle Astrophysics \& Cosmology, Stanford University, 452 Lomita Mall, Stanford, CA 94305, USA}
\author[0000-0001-9504-7386]{Nickolas M. Pingel}
\affiliation{Department of Astronomy, The University of Wisconsin--Madison, 475 N. Charter Street, Madison, WI 53706, USA}
\author[0000-0001-7996-7860]{D. J. Pisano}
\affil{Department of Astronomy, University of Cape Town, South Africa}
\author[0000-0002-2545-1700]{Adam Leroy}
\affiliation{Department of Astronomy, The Ohio State University, 140 West 18th Avenue, Columbus, OH 43210, USA}
\affil{Center for Cosmology and Astroparticle Physics, 191 West Woodruff Avenue, Columbus, OH 43210, USA}
\author[0000-0002-4822-3559]{Smita Mathur}
\affiliation{Department of Astronomy, The Ohio State University, 140 West 18th Avenue, Columbus, OH 43210, USA}
\affil{Center for Cosmology and Astroparticle Physics, 191 West Woodruff Avenue, Columbus, OH 43210, USA}
\author[0000-0002-2155-6054]{George Heald}
\affil{CSIRO Astronomy and Space Science, PO Box 1130, Bentley, WA 6102, Australia}

\begin{abstract}

Diffuse \hi emission in the circumgalactic medium (CGM) of NGC\,891 and NGC\,4565 has been previously shown to trace an inflow along minor axes pointings and to co-rotate with the \hi disk along major axes pointings out to $\approx$100\,kpc \citep{Das2020b,Das2024a}. To obtain a 360\deg view of the inner neutral CGM ($\rm <25 kpc$ for NGC\,891, $\rm <30 kpc$ for NGC\,4565), we perform deep stare observations with the Green Bank Telescope (GBT) along the off-axes, 45$^\circ$ between principal axes, achieving a 5$\sigma$ column density sensitivity of $1.1-1.2\times 10^{17}$\cmsq over a 20 kms$^{-1}$ velocity width. While detecting \hi emission in the inner CGM with single-dish telescopes is common, separating the \textit{true} CGM emission from disk contamination is extremely challenging and has so far been largely unsuccessful. To achieve that, we compare our single-dish detections to deep interferometric maps from the Westerbork Synthesis Radio Telescope (WSRT) HALOGAS survey, and improve upon our previous methods by incorporating velocity offset corrections and channel-wise brightness-temperature scaling. We find that $30-38$\% and $18-28$\% of the emission detected by the GBT cannot be explained by WSRT in NGC\,891 and NGC\,4565, respectively, implying a true CGM detection. There is $4-6\times$($3-7\times$) more \hi along the off-axes than major (minor) axes, nullifying the common assumption of azimuthal symmetry of the neutral CGM. The velocity profile of the diffuse inner CGM suggests a lagged co-rotation with the \hi disk in both galaxies. This exercise illustrates the power of deep observation and careful cross-instrument comparisons to characterize the diffuse \hi in the CGM.  


\end{abstract}

\section{Introduction} \label{sec:intro}
The majority of star formation in a spiral galaxy is confined to the stellar disk within the interstellar medium (ISM). Neutral hydrogen, \hin, is considered to be the raw fuel for star formation. However, the amount of \hi in the ISM is insufficient to sustain star formation on an order of magnitude scale throughout cosmic time \citep{Sancisi2008,Walter2020,Kamphuis2022}. This implies there must be an additional reservoir of \hi gas contributing to the star formation in a galaxy. 

Around the disk of a spiral galaxy and out to the virial radius, a pervasive multiphase gas reservoir known as the circumgalactic medium (CGM) plays an active role in galaxy evolution \citep{Putman2012,Tumlinson2017}. In the CGM, \hi can exist in a range of scales, from high or intermediate velocity clouds and/or as large-scale diffuse low N(\hin) gas. The missing \hi mass needed to explain the SFR in the disk is likely low N(\hin) and in the CGM. However, this regime of low N(\hin) gas in the CGM is largely untapped as it is an immense challenge for both simulations and observations.

Theoretically, there are two main mechanisms of how gas, especially \hin, from the CGM accretes onto the disk: ``hot mode" and ``cold mode" \citep{Keres2005}. Predominantly in less dense environments and in more massive (M$_{200} > 10^{12}M_{\odot}$) and lower-z galaxies (z $\leq$ 3), ``hot mode" accretion involves shock heating the gas to the virial temperature, then a fraction of \hi cools and falls towards the disk in a fragmented but isotropic fashion \citep{Fabian1984}. Alternatively, ``cold mode" accretion dominates in group environments and lower mass and higher-z galaxies, where the \hi gas falls towards the disk in filamentary or sheetlike structures without being shock-heated or significantly interacting with the surrounding hot virialized CGM. The question of ``hot mode" versus ``cold mode"  is still ongoing and is a direct consequence of a) our current inability to observe and simulate the characteristic cold gas scale and b) an unclear understanding of how either accretion mechanism affects the evolution and properties of the host galaxy \citep{McCourt2018,Faucher-Giguere2023}. Detecting strong indicators of either accretion mechanism will be a valuable constraint on the models of galaxy formation, but it has yet to be done. In addition to the two theories of large-scale accretion (from IGM or gas-rich mergers), gas can accrete onto the halo through smaller-scale means, such as through accretion of High Velocity Clouds (HVCs) at rates comparable to their SFR \citep{Lucchini2025}. 

For observers, detecting \hi often comes down to limitations in sensitivity. \hi is traced in two main ways, absorption or emission line studies. In UV, Lyman series and Lyman break absorption studies can achieve as low a N(\hin) of gas as 10$^{12.5}$cm$^{-2}$. Previous Ly$\alpha$ studies have been successful in detecting \hi, N(\hin) $\geq 10^{18}$ cm$^{-2}$ in the Milky Way, and other external galaxies at a redshift of 0.2-0.85 \citep{Richter2017, Borthakur2016,Peroux2022}. Ly$\alpha$ investigations are not ideal when there is no available QSO sightline and/or when Lyman series lines become saturated but undamped at column densities between 10$^{16-18.5}$cm$^{-2}$ (often at low impact parameter, $\sim$ 20 kpc, from the disk). 

Emission studies are done via the 21-cm line corresponding to the spin-flip transition of \hin. Emission can detect N(\hin) in the saturated and undamped regime of Ly$\alpha$ with no requirement of a background source. 21 cm emission presents the best opportunity for us to probe the 10$^{16-18.5}$cm$^{-2}$ N(\hin) regime with no prior density model or background quasar. Detecting the 21 cm line requires the use of either single-dish radio telescopes or an interferometer. 

Single-dish telescopes, such as the Green Bank Telescope (GBT), probe all angular scales, due to their full $uv$ coverage, and achieve high surface brightness sensitivity. The Green Bank Telescope has an unblocked aperture, high aperture efficiency, and high surface brightness sensitivity (T$_{B} \leq 20$ K). GBT has low sidelobes and can achieve an angular resolution of 9.1$'$. As a result, GBT is an ideal single-dish candidate to look for low N(\hin) gas over large angular scales.

Interferometers are networks of smaller single-dish telescopes that together image one source at high angular resolution but with a low surface brightness sensitivity. Interferometers cannot probe larger angular scales due to the $uv$ spacing problem, with their largest resolvable angular scale being inversely proportional to the distance between the farthest telescopes; a tighter configuration allows us to probe larger angular scales. 
Interferometric surveys, including Hydrogen Accretion in LOcal GAlaxies Survey \citep[\href{https://www.astron.nl/halogas/data.php}{HALOGAS},][]{Heald2011}, and MeerKAT HI Observations of Nearby Galactic Objects; Observing Southern Emitters \citep[\href{https://mhongoose.astron.nl/sample.html}{MHONGOOSE},][]{Sorgho2019,deBlok2024} survey have contributed a wealth of knowledge regarding the \hi distribution of nearby galaxies at high angular resolution through mappings.

There are two observational methods employing both interferometers and single-dish: mapping and deep stare. Mapping is when the instrument slews across multiple points with a short integration time. Mapping achieves a shallower N(\hin) as a result of the short integration time; however, it does provide the spatial distribution of the emission. The Deep Stare technique involves staring at one pointing for an extended amount of time to achieve a surface brightness sensitivity at that particular pointing, although at the cost of lacking detailed spatial distribution of any detected emission. When compared at the same spatial and velocity resolution, single-dish telescopes will detect all the emission detected by interferometers, with the addition of large angular scale, low column-density gas. In a crude sense, when observing the same object, the difference between a single dish and an interferometer corresponds to the diffuse gas. Thus, to isolate only the low column density gas, we require the use of both radio instruments: single-dish and interferometers.

\cite{Pisano2014},\cite{deBlok2014}, mapped the CGM of NGC\,2997, NGC\,6946, and NGC\,2403 using GBT at a 5$\sigma$ sensitivity of $10^{18}$ \cmsq over a 20 kms$^{-1}$ channel revealed the presence of excess low column density gas in GBT in all three galaxies and the presence of low column density clouds and filaments in NGC\,2403, NGC\,6946, respectively. \cite{Howk2017} examined 48 sightlines in the CGM of M\,31 using GBT at 5$\sigma$ sensitivity of $10^{17.5}$ \cmsq at an impact parameter of 23--340\,kpc. No \hi emission was detected, except for two sightlines passing through the Magellanic Stream, implying that either M\,31 is an exception or successful \hi emission detection requires larger sky coverage. GBT observations of four local edge-on galaxies, NGC\,891, NGC\,925, NGC\,4414, and NGC\,4565, were compared against HALOGAS, at the same velocity and spatial resolution \citep[hereafter \citetalias{Pingel2018}]{Pingel2018}. No significant excess emission was detected in GBT compared to HALOGAS, implying that either these galaxies are also exceptions or the diffuse \hi gas is below their 5$\sigma$ GBT detection limit of 0.9–1.4 $\times$ 10$^{18}$ \cmsq over a 20 km$^{-1}$ channel, meaning deeper observations are needed. Observations with the Five-hundred-meter Aperture Spherical Radio Telescope (FAST) observe galaxies from the THINGS survey (observed by the VLA), and do not find evidence of gas at column densities lower than $10^{17.7}$ cm$^{-2}$ \citep{Wang2024}. The non-detections imply that if these galaxies host \hin, it must have N(\hin) $< 10^{17.7}$ cm$^{-2}$. 

\begin{deluxetable*}{cccccccccc}
	\centering
   \tablecaption{GBT Data Reduction Details} \label{tb:datared}
   \tablehead{\colhead{Pointing$^{(a)}$} & \colhead{RA} & \colhead{DEC} &\colhead{t$\rm{_{eff}}$$^{(b)}$}& \colhead{T$\mathrm{_{sys}}$} & \colhead{ $\rm{\sigma_{T_{B}}}^{(c)}$ } & 
   \colhead{d$\mathrm{_{\perp}}^{(d)}$} 
    & \colhead{$\Delta$v$\mathrm{_{emit}}$$^{(e)}$} 
     & \colhead{$\Delta$v$\mathrm{_{base}}$$^{(f)}$}
     & \colhead{n$\mathrm{_{fit}}$$^{(g)}$}}
\startdata
 & (J2000) & (J2000) & (hrs) & (K) & (mK) & (kpc) & $\mathrm{(km s^{-1})}$  & $\mathrm{(km s^{-1})}$  &  \\
 \midrule
\multicolumn{10}{c}{\textbf{NGC\,891 (2$^{h}$22$^{m}$33.6$^{s}$, 42$^{\circ}$20$^{'}$58$^{''}$)}: $\rm \delta v^{(h)} = 2.5\,km\,s^{-1}$}\\
 \midrule
NO\,1E & 2$^{h}$23$^{m}$18.06$^{s}$ & +42$^{\circ}$24$^{'}$36$^{''}$ & 1.12 & 18.35 & 2.21 & 24.0 & (-270, 233)& (-450, 650)  & 0 \\
NO\,1W & 2$^{h}$22$^{m}$15.34$^{s}$  & +42$^{\circ}$29$^{'}$25$^{''}$ & 1.12 & 18.36 & 2.21 & 24.3 &  (-265, 233) & (-450, 650) & 0  \\
SO\,1E & 2$^{h}$22$^{m}$53.18$^{s}$  & +42$^{\circ}$12$^{'}$33$^{''}$ & 1.08 & 18.36 & 2.33 & 24.5 &  (-245, 281)& (-400, 600) & 1  \\
SO\,1W & 2$^{h}$21$^{m}$48.66$^{s}$  & +42$^{\circ}$17$^{'}$29$^{''}$ & 1.08 & 18.35 & 2.16 & 24.1 &  (-255, 271)& (-450, 650) & 1 \\
 \midrule
 \multicolumn{10}{c}{\textbf{NGC\,4565 (12$^{h}$36$^{m}$20.9$^{s}$, 25$^{\circ}$59$^{'}$23$^{''}$)}: $\rm \delta v^{(h)} = 2.5\,km\,s^{-1}$}\\
 \midrule
NO\,1E & 12$^{h}$36$^{m}$24.0$^{s}$  & +26$^{\circ}$ 08$^{'}$44$^{''}$ & 1.12 & 18.01 & 7.31 & 29.5 & (-113, 282) & (-1000, 625) & 3  \\
NO\,1W & 12$^{h}$35$^{m}$39.4$^{s}$  & +26$^{\circ}$00$^{'}$10$^{''}$ & 1.06 & 18.01 & 2.57 & 29.4 &  (-205, 280)& (-1000, 625) & 3 \\
SO\,1E & 12$^{h}$37$^{m}$02.6$^{s}$  & +25$^{\circ}$58$^{'}$42$^{''}$ & 1.04 & 17.97 & 2.81 & 29.5& (-292, 139)& (-1000, 625) & 3 \\
SO\,1W & 12$^{h}$36$^{m}$18.3$^{s}$  & +25$^{\circ}$50$^{'}$01$^{''}$ & 1.04 & 17.96 & 2.58 & 29.5 &  (-294, 209)& (-1000, 625) & 3 \\
\enddata
{\small $(a)$ ``NO" and ``SO" denote offsets to higher and lower declination than the center of the target galaxy. ``E" and ``W" denote offsets to higher and lower right ascension than the center of the target galaxy.} 

{$(b)$ Effective exposure time t$_{\rm eff}$ = t$_{\rm on}\times$t$_{\rm off}$/(t$_{\rm on}$+t$_{\rm off}$). t$_{\rm on}$ and t$_{\rm off}$ are the integration times toward the on-source and the off-source pointings, respectively. To maximize t$_{\rm eff}$, we maintain t$_{\rm on}$ and t$_{\rm off}$ to be the same.} 

{$(c)$ The RMS noise from the signal-free region of the baseline-subtracted spectrum.} 

{$(d)$ The transverse distances of the pointings from the center of the target galaxy, adopting a distance of 9.2\,Mpc and 10.8\,Mpc for NGC\,891 and NGC\,4565 from \cite{Heald2012}}. 

{$(e)$ The velocity range of detected \hi emission. }

{$(f)$ The velocity range of the baseline.}

{$(g)$ The order of the polynomial used to best fit the baseline.}

{$(h)$ Velocity resolution of the spectrum from the baseline subtraction process; please refer to Appendix\,A of \citetalias{Das2024a} for a detailed description of the baseline subtraction routine.}
\end{deluxetable*}

The MHONGOOSE survey, which examined 30 local galaxies using the MeerKAT radio interferometer to study the disc halo interface of these galaxies, achieves an unprecedented \hi column density sensitivity  (3 $\sigma$ over 50 km s$^{-1}$) of $5\times 10^{17}$ cm$^{-2}$ at 90 '' resolution to $4\times 10^{19}$ cm$^{-2}$ at 7'' \citep{deBlok2024}. 
Recent work shows several MHOONGOOSE galaxies host extraplanar low-column-density gas (a few times 10$^{18}$cm$^{-2}$), low surface brightness companion galaxies, extraplanar filamentary structures, and extended co-rotating \hi disks \citep{Maccagni2024,Namumba2025,Kurapati2025,Veronese2025,Healy2024}.
Examining 18 local MHONGOOSE galaxies with GBT, \cite{Sardone2021} found that in 15 galaxies \hi profiles do not truncate at large impact parameters, indirectly suggesting the presence of diffuse extraplanar \hin. 
Although interferometers, such as MeerKAT, have made significant improvements in the form of higher surface brightness sensitivity and better angular resolution, they are still fundamentally limited by the short spacing problem. Even with world-class instruments, detecting the lowest column density gas at high angular scales would still require an egregious use of telescope time. The comparison with a single dish is therefore still essential in detecting the low N(\hin) gas around individual galaxies. Still, there is a mixed bag of successes in many studies in detecting diffuse gas in the CGM of spiral galaxies. This perpetuates the question of the form of the diffuse gas; is it smeared out clouds? An extended \hi disk? To determine the morphology at the disk-halo interface, we require an azimuthal view of the \hi.

\begin{figure*}
\centering
    \includegraphics[trim=65 10 115 5, clip,width=0.49\linewidth]{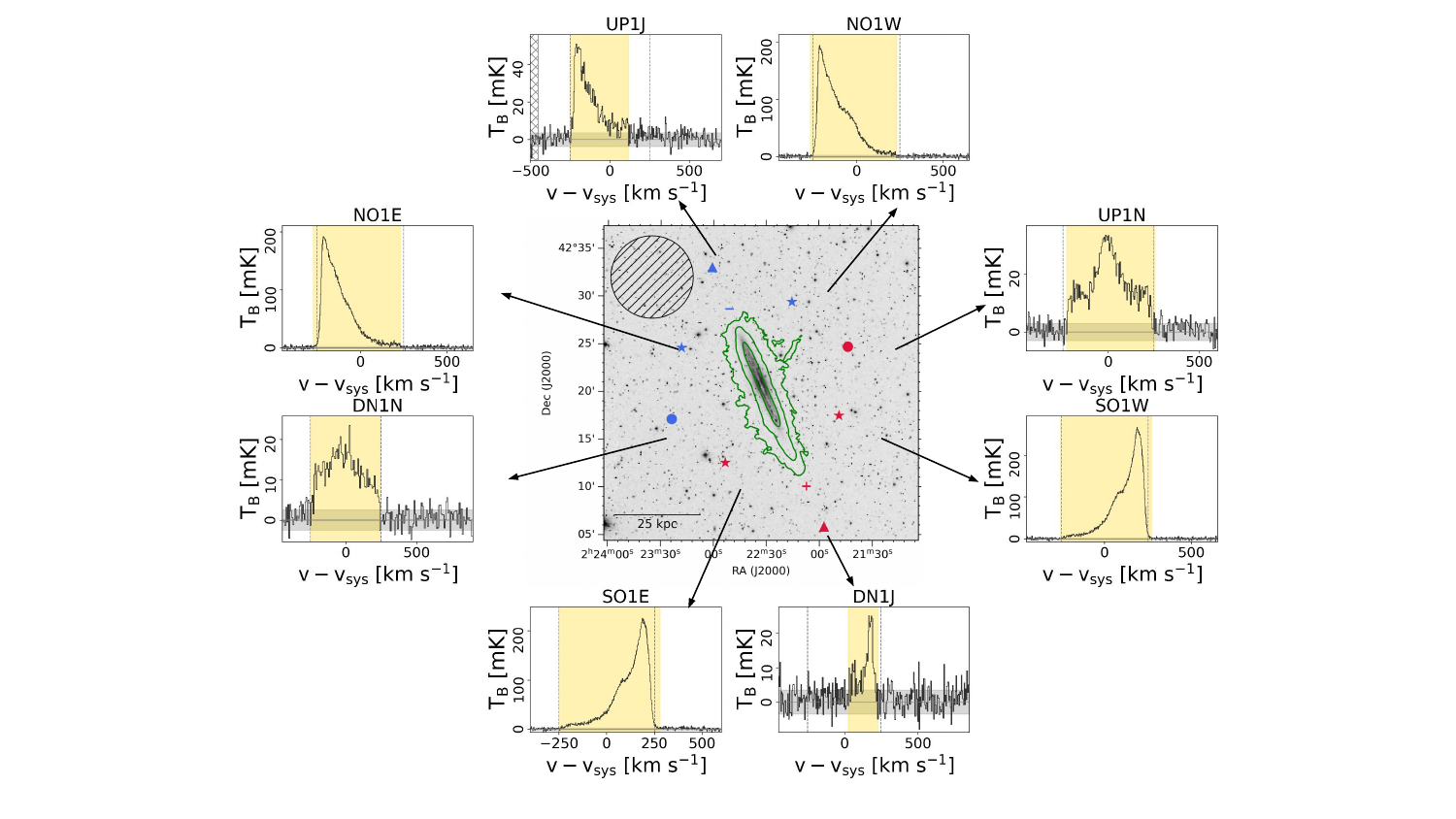}
    \includegraphics[trim=95 10 90 5, clip,width=0.49\linewidth]{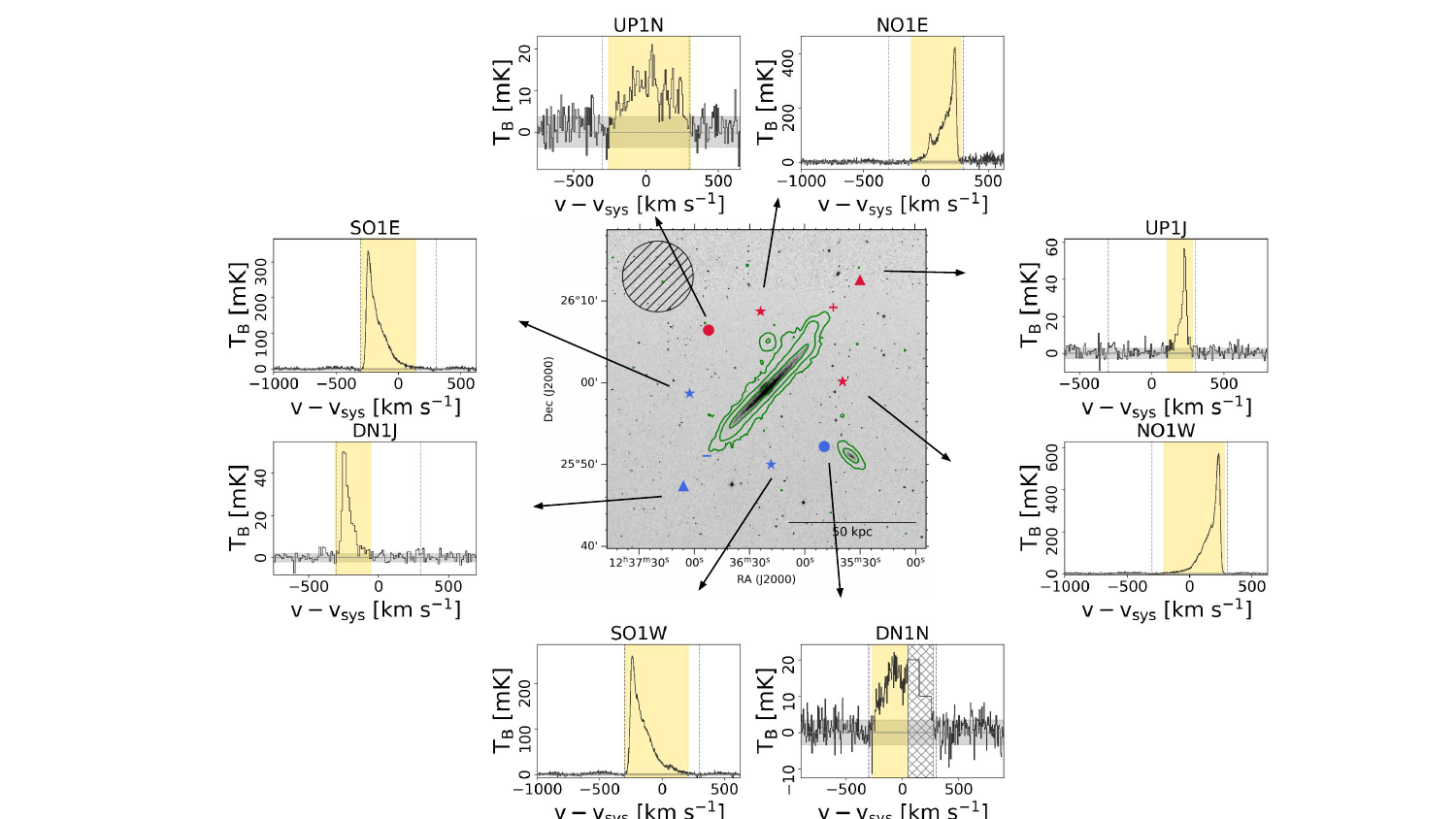}
    \caption{21-cm GBT spectra of the inner CGM of NGC\,891 (left) and NGC\,4565 (right) are shown around the Digitized Sky Survey (DSS) image. Overlayed in green on the DSS images are the 21-cm contours from the HALOGAS survey, with the lowest contours at 10$^{19}$ cm$^{-2}$. The triangles, circles, and stars correspond to the positions of major axes, minor axes, and off-axis pointings from GBT projects 20B-360 \citepalias{Das2024a}, 15B-257 \citepalias{Das2020b}, and 21B-324 (this work); the red and blue colors denote the red/blueshifted velocity in the rest frame of the galaxy. Note that all CGM pointings are one beam away from the edge of the \hi disk, making the off-axes and minor axes pointings approximately equidistant from the galaxy, and the major axes pointings further away. In the spectral plots, vertical gray dashed lines correspond to the maximum line-of-sight velocity of the \hi disk emission, and the horizontal solid gray line corresponds to T$_{B}=$0. The vertical regions highlighted in yellow correspond to the velocity width of the detected emission. The full $\Delta$v$\mathrm{_{base}}$ (see Table \ref{tb:datared}) baselines are not plotted for ease of viewing. The \textit{true} CGM emission is hidden under the dominant disk contamination in these spectra, which we assess in the following sections.}
    \label{fig:baseline}
\end{figure*}

The predecessors of this work, \cite{Das2020b,Das2024a}, hereby referred to as \citetalias{Das2020b,Das2024a}, observed two galaxies, NGC\,891 and NGC\,4565, using a deep stare with the GBT.  
These observations stepped along the minor and major axes of each galaxy out to 3 GBT FWHM beam sizes away from the target, integrated 3--4 hours each pointing, and achieved an unprecedented 5$\sigma$ column density sensitivity of $5\times 10^{16}$ \cmsq over a 20 km\,s$^{-1}$ velocity width. In nearly all pointings, diffuse \hi was detected. In this paper, we complete the 360$^\circ$ view of the diffuse \hi in the inner CGM of NGC\,891 and NGC\,4565 by examining four off-axis pointings, 45$^\circ$ between the principal axes, using similar observing and analysis methods.

\cite{Yang2025} observed several nearby edge-on spiral galaxies with the FAST, including NGC\,891 and NGC\,4565, at a 3$\sigma$ N(\hin) sensitivity of $5\times 10^{17}$ cm$^{-2}$ at 20 km s$^{-1}$ velocity resolution, and compared it with different available interferometer maps, specifically using HALOGAS for NGC\,891 and NGC\,4565. They detected excess, diffuse \hi emission in all galaxies along the minor axes with varying radial profiles and clear extraplanar features extending 20--50\,kpc from the disk, with the extraplanar diffuse CGM contributing 5\% to 10\% of the total \hi mass.

\section{GBT Observations} \label{sec:style}

\subsection{Sample Selection}

NGC\,891 and NGC\,4565 are pilot targets for many reasons. To study the azimuthal distribution of \hi in the CGM, we need to observe along both major, minor, and off-axes, limiting us to edge-on galaxies. To separate diffuse gas from ``clumpy" emissions, we require access to existing interferometric surveys. Motivated by \citetalias{Pingel2018,Das2020b,Das2024a}, and  \cite{Sardone2021}, interferometric maps can be convolved with the GBT beam and compared to GBT spectra. NGC\,891 and NGC\,4565 are both observed in the HALOGAS survey, which is a deep (10$^{19}$ \cmsq) map of each target galaxy. Additionally, we require the galaxies to be not too close to ensure that we sample a decent spatial region of the CGM. 
We also require the systemic velocities to be sufficiently different from the Milky Way's emission to allow for clear baseline subtraction.

\subsection{Observational details}

The sky position of each pointing was determined based on previous examinations of their \hi disks. In GBT projects 15B-257 (PI: Leroy), 20B-360 (PI: Das), and 21B-324 (PI: Das), \citetalias{Das2020b,Das2024a} probed along the major and minor axes. The minor axis pointings were closer to the galaxy center than the major axis pointings because of the \hi disk truncation at different impact parameters (see Fig.\,1). The off-axis pointings (GBT project 21B-324) are located at 45$^\circ$ azimuthal angle between the minor (referred to as N) and major axis (referred to as J), 1 FWHM GBT beam away from the edge of the \hi disk as measured by interferometric surveys (see Figure \ref{fig:baseline}. Along the off-axes, we label high/low declination pointings as north (NO)/south (SO) and high/low right ascension as east (E)/west (W), yielding pointings NO\,1E, SO\,1E, SO\,1W, and NO\,1W (see Fig.\,\ref{fig:baseline}). We ensure there is no known \hi emitter in the line of sight of our pointings. 

Our observations entail 20 position-switching GBT sessions in two polarizations (XX and YY), using the L-band receiver and Versatile GBT Astronomical Spectrometer (VEGAS; bandwidth $=$ 23.5 MHz) as the backend. Off-sources are chosen to be 1.5\deg, $\sim$ 200 and 460 kpc for NGC\,891 and NGC\,4565, away from the disk at a similar declination to the on-source pointings to avoid additional noise arising from overhead time in changing the declination. We select the flux calibration sources to be 3C\,123/3C\,48 and 3C\,286 for NGC\,891 and NGC\,4565, respectively. 

\section{Data reduction and analysis}\label{sec:datared}

We use a GBTIDL-based reduction pipeline developed in \citetalias{Das2020b,Das2024a} to reduce the position-switched spectra. From the pipeline, we extract the effective integration time (\teff\!), average system temperature (\Tsys\!), and spectral weight t$_{\rm eff}$/T$_{\rm sys}^2$, for both polarizations in each session. We visually identify residual radio frequency interference (RFI) channels and then linearly interpolate these channels using the adjacent channels. For each pointing, we stack all sessions in each polarization, weighted by the spectral weight of each session, to obtain mean spectra. We average the spectra from both polarizations into one spectrum. We report the average \Tsys and \teff over multiple sessions in Table\,\ref{tb:datared}. We subtract the systemic velocity of the galaxy (530 and 1230 km s$^ {- 1} $ for NGC\,891 and NGC\,4565), so all spectra are in the reference frame of the respective galaxy. 

We fit and subtract a baseline from each spectrum using an iterative polynomial fitting method outlined in \citetalias{Das2024a}. We allow the polynomial order to vary from 0--3\footnote{Order 3 is used only in NGC\,4565, which had a wavy baseline}. \citetalias{Das2024a} considered velocity resolution to be 5-20 km\,s$^{-1}$, smoothing out for farther-away pointings, and sacrificing detailed kinematic information to detect the integrated signal. In the off-axis spectra, baselines are relatively simple, and the S/N ranges from 280 to 380. With such strong signals, we can preserve kinematic information and restrict our velocity resolution to 2.5 km\,s$^{-1}$. When comparing 2.5 and 5 km\,s$^{-1}$ velocity resolution, we found no significant improvement. We plot these baseline-subtracted spectra for NGC\,891 (left) and NGC\,4565 (right) in Figure \ref{fig:baseline}.

We calculate the integrated intensity in GBT by integrating the T$_{\rm{B}}$ over $\Delta$v$_{\rm emit}$ (see Table \ref{tb:datared}). Assuming the emission is optically thin and the number density of the gas is well above the critical density of \hi for collisions with electrons \citep[7$\times$10$^{-6}$ cm$^{-3}$ at 10$^4$ K;][]{Draine2011}, we convert integrated T$_{\rm{B}}$ to N(\hin) to obtain N(\hin)$\rm _{GBT}$. In Table \ref{tb:deriv}, we quote the two uncertainties of N(\hin)$\rm _{GBT}$: statistical and systematic. The statistical uncertainty is calculated by multiplying the RMS noise (from the signal-free part of the baseline-subtracted spectrum) by the square root of the product of $\rm \delta v$ and $\rm \Delta v_{emit}$ (see Table \ref{tb:datared}, described in greater detail in Appendix\,A in \citetalias{Das2024a}. We use the RMS of the baseline as the statistical uncertainty on the GBT measurements. The systematic uncertainty comes from the variation in flux calibration across observing sessions. 


\begin{deluxetable*}{ccccccccc}
\centering
\tablecaption{Derived Quantities} \label{tb:deriv}
\tablehead{
\colhead{Pointing} & \colhead{N(\hin)$\mathrm{_{GBT}}^{(a)}$} & \colhead{$\mathrm{v_{GBT}}$} &\colhead{M(\hin)$\mathrm{_{GBT}}$}& \colhead{N(\hin)$\mathrm{_{WSRT}}^{(b)}$} &  \colhead{N(\hin)$\mathrm{_{CGM}}^{(c)}$} & \colhead{v$\mathrm{_{CGM}}$} & \colhead{M(\hin)$\mathrm{_{CGM}}$} & \colhead{GBT$_{\mathrm{\% excess}^{(d)}}$}\\
& $\mathrm{(\times10^{19} cm^{-2})}$ & $\mathrm{(km s^{-1})}$ & ($\times10^8$ M$_\odot$) & $\mathrm{(\times10^{19} cm^{-2})}$ &  $\mathrm{(\times10^{19} cm^{-2})}$ & $\mathrm{(km s^{-1})}$ & ($\times10^7$ M$_{\odot}$)  
}  
\colnumbers
\startdata
\multicolumn{9}{c}{\textbf{NGC\,891: $\rm {v_{offset}}^{(e)} = 0\,km\,s^{-1}$}}\\
\midrule
NO\,1E & 5.03$\pm$0.02$\pm$0.09 & -122$\pm$1 & 1.70$\pm$0.03 & 3.30$\pm$0.22& 1.73$\pm$0.24 & -71$\pm$26 & 5.82$\pm$0.79 & 34$\pm$7  \\
NO\,1W & 5.39$\pm$0.02$\pm$0.10 & -113$\pm$1 & 1.82$\pm$0.03 & 3.33$\pm$0.25 & 2.06$\pm$0.27 & -63$\pm$24 & 6.95$\pm$0.90 & 38 $\pm$ 7 \\
SO\,1E & 5.83$\pm$0.02$\pm$0.10 & 120$\pm$0 & 1.96$\pm$0.03 & 3.81$\pm$0.32 & 2.02$\pm$0.34 & 67$\pm$10 & 6.79$\pm$1.15 & 35 $\pm$8 \\
SO\,1W & 6.74$\pm$0.02$\pm$0.12 & 122$\pm$0 & 2.27$\pm$0.04 & 4.71$\pm$0.40 & 2.03$\pm$0.42 & 57$\pm$15 & 6.84$\pm$1.41 & 30 $\pm$ 9 \\
\midrule
\multicolumn{9}{c}{\textbf{NGC\,4565: $\rm {v_{offset}}^{(d)} = 7\,km\,s^{-1}$}} \\
\midrule
NO\,1E & 7.44$\pm$0.04$\pm$0.02 & 161$\pm$0 & 3.46$\pm$0.02 & 5.56$\pm$0.42 & 1.88$\pm$0.42 & 100$\pm$15 & 8.73$\pm$1.95 & 25 $\pm$ 8 \\
NO\,1W & 9.21$\pm$0.02$\pm$0.02 & 167$\pm$0 & 4.28$\pm$0.01 &  7.34$\pm$0.62 & 1.87$\pm$0.62 & 72$\pm$35 & 8.68$\pm$2.89 & 20$\pm$10 \\
SO\,1E & 6.59$\pm$0.02$\pm$0.02& -172$\pm$1 & 3.06$\pm$0.01 & 4.91$\pm$0.42 & 1.69$\pm$0.42& -131$\pm$74 & 7.84$\pm$1.93 & 26 $\pm$ 9 \\
SO\,1W & 5.95$\pm$0.02$\pm$0.01 & -156$\pm$1 & 2.77$\pm$0.01 & 4.56$\pm$0.39 & 1.39$\pm$0.39& -108$\pm$74 & 6.45$\pm$1.80 & 23 $\pm$9  \\
\bottomrule
\enddata
{$(a)$ N(\hin) detected by GBT over $\Delta$v$\mathrm{_{emit}}$ followed by the statistical uncertainty and then multiplicative systematic uncertainty from variation in flux calibration across GBT observing sessions.}

{$(b)$ N\hi detected by masked WSRT data using the method described in \citetalias{Das2024a}, over $\Delta$v$\mathrm{_{emit}}$. Uncertainties are the quadrature sum of statistical uncertainty (of order $\sim$ 0.0001) and systematic uncertainty (dominant source of uncertainty of order $\sim$ 0.01). Where the systematic uncertainty is the quadrature sum of the multiplicative systematic uncertainty from azimuthal beam asymmetry and the additive systematic uncertainty in the masking threshold.}

{$(c)$ Diffuse (excess) \hi detected by the GBT compared to WSRT in the CGM. The error is the quadrature sum of statistical uncertainties and systematic uncertainties in N(\hin) for both GBT and WSRT.}

{$(d)$ Percent observed by GBT that cannot be explained by WSRT. This excess is attributed to a detection of the diffuse CGM.} 

{$(e)$ Velocity correction, shown in cartoon Figure \ref{fig:appdx_velosig}, applied to the WSRT spectrum to line up the rising or falling edge with GBT. The value, if non-zero, is applied to both the central pointing (used to calculate the channel-wise scaling factor) and CGM pointings. Detailed methodology of the velocity edge correction is described in Appendix \ref{appdx:velocorrect}. }

\end{deluxetable*}

\begin{figure*}
\centering
    \includegraphics[trim=0 25 300 0, clip,width=0.29\textwidth]{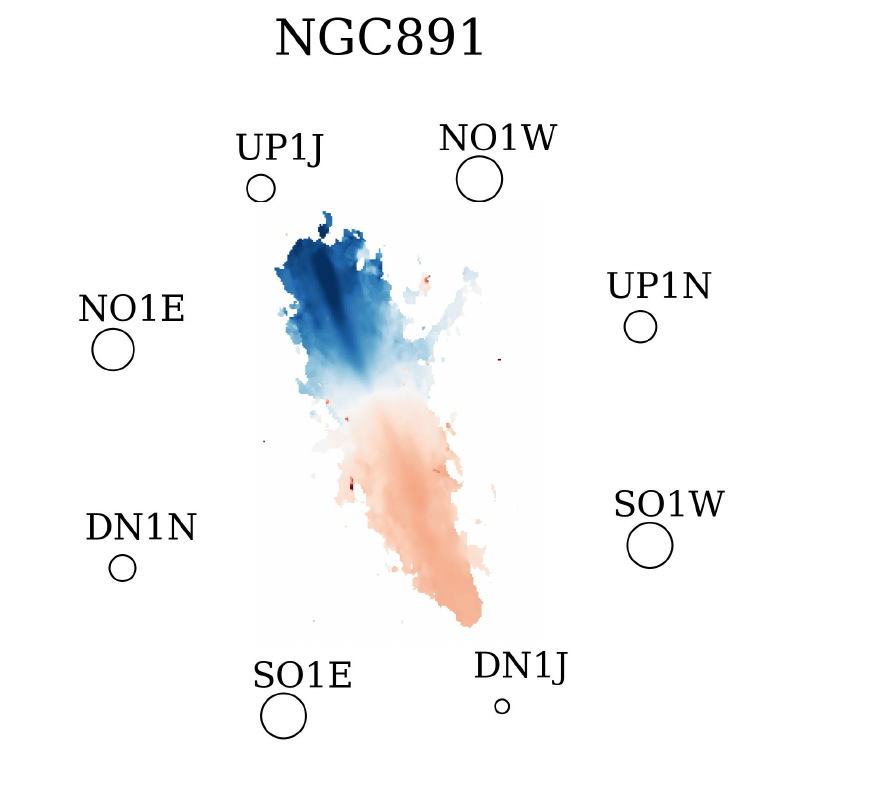}
    \includegraphics[width=0.67\textwidth]{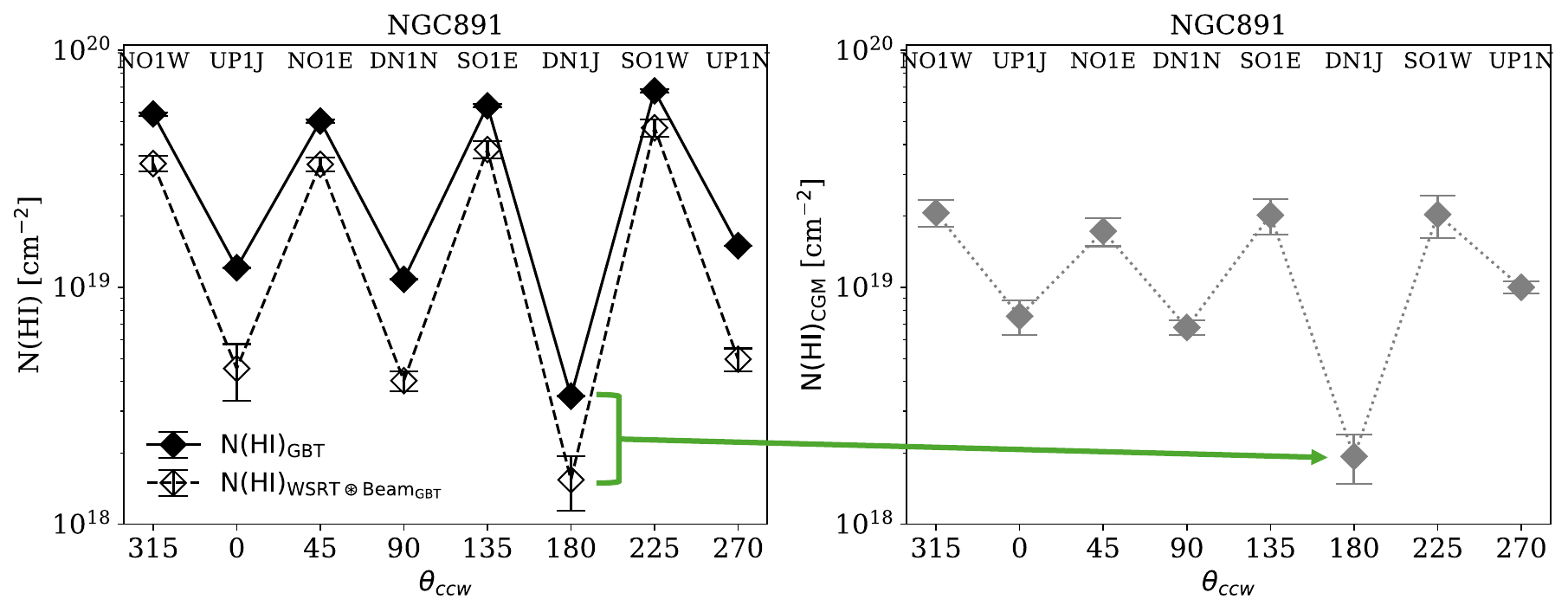}
    \includegraphics[trim=0 25 300 0, clip,width=0.29\textwidth]{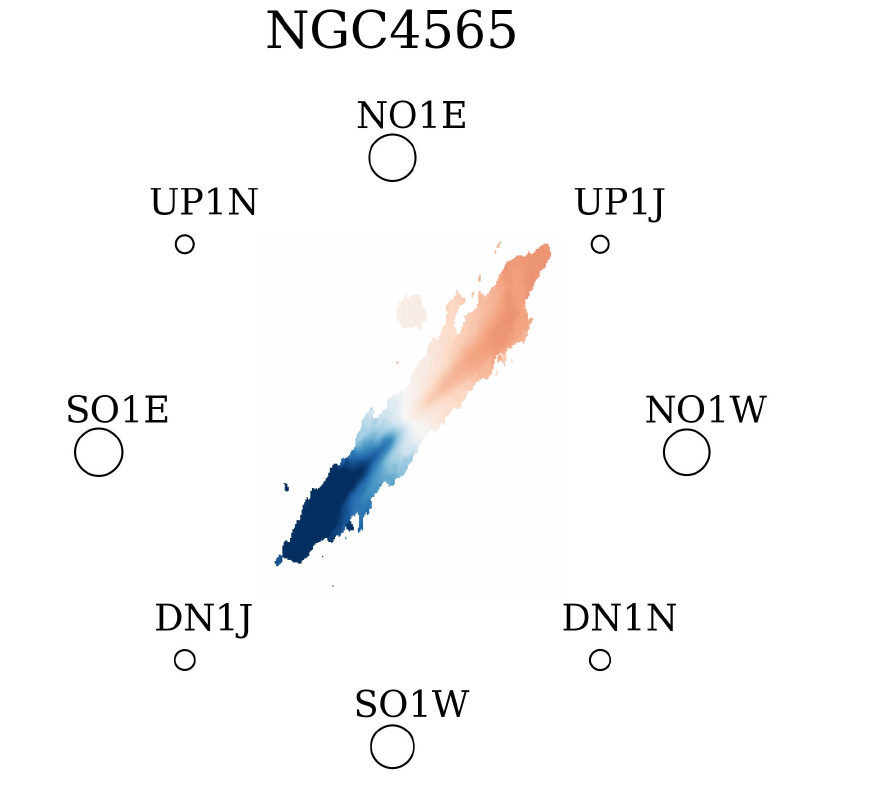}
    \includegraphics[width=0.67\textwidth]{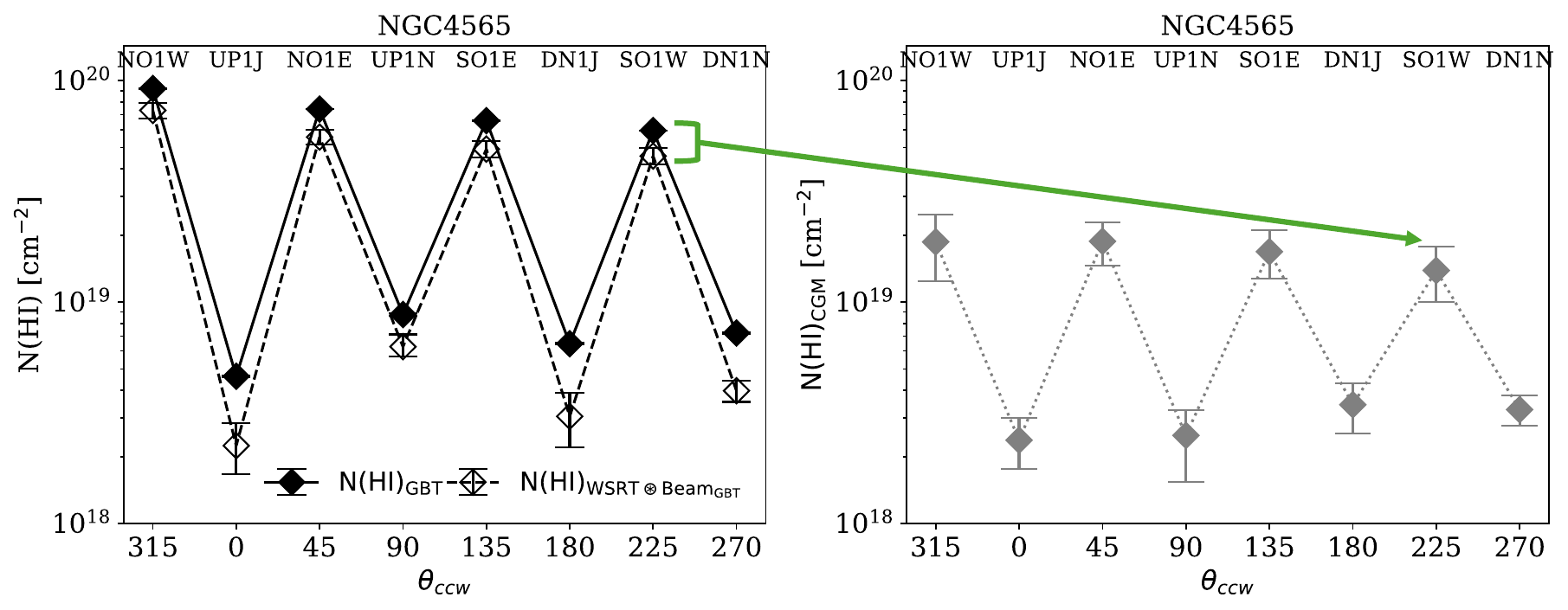}
    \caption{\textbf{Left:} Position of pointings around NGC\,891 (top) and NGC\,4565 (bottom), with larger circles corresponding to pointings with higher N(\hin). The pointings surround the HALOGAS moment-one map, which is color-coded to indicate the blue/red-shifted velocities of the \hi disk. \textbf{Middle}: N(\hin) from the GBT (filled symbols) and WSRT convolved with the GBT beam (unfilled symbols), starting at NO\,1W and following other pointings counter-clockwise. The azimuthal angle, $\theta_{\mathrm{CCW}} = 0^\circ$ corresponds to UP\,1J; see detailed descriptions in \S\ref{sec:results}.  Clearly, GBT detects more \hi than what the WSRT does at each pointing, revealing the detection of the \textit{true} CGM emission (\textbf{right} panel). The GBT data (\textbf{middle} panel) is dominated by the disk contamination measured by the WSRT data, but as the fractional contamination is not uniform across pointings, the residual circumgalactic \hi (\textbf{right} panel) features an azimuthal variation which differs across pointings; see Section \ref{sec:sysortrue} for a thorough investigation of the azimuthal variation.}
    \label{fig:891}
\end{figure*} 

\subsection{Assessing disk contamination with interferometers}\label{sec:comp_interfer}

Because the CGM pointings are one GBT beam-size away from the edge of \hi disk, the measurements from GBT in Figure \ref{fig:891} are heavily dominated by contamination from the \hi disk. To separate the true CGM emission from the disk contamination,  we require the rigorous comparison of our single-dish (GBT) data to interferometric (WSRT HALOGAS) data as done in \citetalias{Pingel2018,Das2020b,Das2024a}.

The WSRT spectrum serves as the characterization of stray light from the \hi disk contaminating the GBT data at our CGM pointings. If the GBT and WSRT measurements are consistent with each other, it would imply that there is no detectable CGM emission. If there is any excess in GBT compared to WSRT, it would indicate the presence of diffuse CGM missed by WSRT due to its shallower sensitivity and limited uv-spacing.

To subtract disk contamination assessed by WSRT from the GBT data, we must first convert the WSRT cubes to the same angular resolution as GBT via convolution (see \S\ref{sec:convolve}) and then apply corrections necessary to account for instrumental differences between GBT and WSRT (see \S\ref{sec:crossinstrum}). We emphasize that any final result on the CGM is not based solely on GBT or WSRT, but rather on the excess detected in the GBT data after disk contamination (WSRT) is removed. In the following two subsections, we discuss in detail how we have assessed disk contamination using WSRT data to separate the subdominant CGM emission in the GBT data, including all known statistical and systematic uncertainties.

\subsubsection{Convolution of the interferometric cubes}
\label{sec:convolve} 

To reconstruct WSRT data at the same angular resolution and velocity space as our GBT data, we follow the steps outlined in \citetalias{Das2020b,Das2024a}, which we summarize below for clarity.
\begin{enumerate}
    \item \textbf{Noise Estimation:} We estimate the global noise, $\sigma$, by fitting a Gaussian to the histogram of pixels with negative values beyond 20$'$. 
    \item \textbf{Primary beam correction:} To correct for the primary beam response of the HALOGAS cubes, we use the MIRIAD task \textit{pbcor}.
    \item \textbf{Masking:} To avoid the negative bowl and any spurious emission in the WSRT cube, we mask it at 4, 5, and 6 $\sigma$ levels. The 5$\sigma$ masked cube is used to calculate the best estimate of the WSRT spectra (N(\hin)$_{\rm{WSRT}}$ in Table \ref{tb:deriv}; see Appendix B in \citetalias{Das2024a} for a detailed discussion on selecting the best S/N threshold).  N(\hin)$_{\rm{WSRT}}$ is followed by the quadrature sum of the statistical and systematic uncertainty. The statistical uncertainty in WSRT comes from the global $\sigma$ weighted by the GBT beam response (See \citetalias{Das2020b} Section 3.1). The systematic uncertainty is the square root quadrature sum of the difference between masking levels 5$\sigma$ and 4/6$\sigma$ (lower/upper), and the uncertainty from the azimuthal variation of the GBT beam (described in detail in \S3.1 of \citetalias{Das2020b}). For a discussion on the robustness of this masking technique, see Appendix \ref{appdx:crossinstrument} ``Method 1".     
    \item \textbf{Convolution:} We convolve the masked cubes with the circularized GBT beam centered on our CGM pointings (same as the method described by \citetalias{Pingel2018}). 
    \end{enumerate}

The resultant masked and convolved WSRT spectra represent what GBT would have seen if GBT observed only disk contamination. Up to this point, we have converted WSRT to be within the same angular space as GBT. In the next section, we discuss additional necessary corrections made to the masked convolved WSRT spectrum to account for cross-instrument differences. 

\subsubsection{Correcting for cross-instrument effects}

\label{sec:crossinstrum}
We observe the \hi disk of both galaxies (the central pointing) at the beginning of each GBT observing session as a reference. We expect the \hi emission from the disk to be comparable between GBT and WSRT, but due to calibration uncertainties across both instruments, it is often different. This discrepancy, if not corrected, translates to the measurements in the CGM pointings as well, leading to incorrect assessment of disk contamination, resulting in incorrect estimation of the \textit{true} CGM emission. In \citetalias{Das2024a}, we introduced an integrated scaling factor, $f_{\rm{scalar}}$, to correct for this. In this paper, we improve that analysis by introducing a channel-wise scaling factor, $f_{\rm{channel}}$, rather than a single channel-averaged value, $f$. We plot the comparison of $f_{\rm{scalar}}$, $f_{\rm{channel}}$, and their effects when applied to the central WSRT spectra.  This preserves velocity information and channel-wise variation of $f$ and is discussed at length in Appendix\,\ref{appdx:Tbscaling}. To calculate $f_{\rm{channel}}$, we apply a pointing-specific correction of the velocity edges to ensure the GBT and WSRT spectra are perfectly aligned in the velocity space (discussed at length in Appendix\, \ref{appdx:velocorrect}). A misalignment in the velocity edges would significantly over- or underestimate $f_{\rm{channel}}$, which would manifest as over- or underestimated N(\hin)$\rm_{WSRT}$. We tested these new steps on the principal axes pointings as well to verify that the results do not change. We include the re-analyzed inner CGM pointings of \citetalias{Das2020b,Das2024a} in Figure \ref{fig:891} and Figure \ref{fig:nhi_v_cgm}.  


\subsubsection{Extracting Physical Quantities}

We calculate the integrated intensity in WSRT by integrating the 5$\sigma$-masked channel-wise scaled T$_{\rm{B}}$ over $\Delta$v$_{\rm emit}$ (see Table \ref{tb:datared}). We convert the integrated intensity to column density, N(\hin)$_{\rm{WSRT}}$. 

We calculate the NHI$\rm_{CGM}$ by subtracting NHI$_{\rm WSRT}$ from NHI$_{\rm GBT}$. We calculate the mass of \hi in the CGM, where m$\rm_p$ is the mass of the proton and dA is the area of the main GBT beam:
\begin{subequations}
\begin{equation}\label{eq:mass}
  {\rm  M(HI)_{CGM}} =  \int m_p {\rm N(HI)_{CGM}} \rm dA  
\end{equation}
We obtain the average line-of-sight velocity of the \hi emission in the CGM, $\rm \langle v_{CGM} \rangle$, using equation \ref{eq:vavg}. 
\begin{equation}\label{eq:vavg}
   \rm \langle v_{CGM} \rangle = 
   \frac{\int \left(T_{B,G} - T_{B,W}\circledast Beam_{G}\right) v_{G} \, dv_{G}}
        {\int \left(T_{B,G} - T_{B,W}\circledast Beam_{G}\right) \, dv_{G}}
\end{equation}
\end{subequations}

Here, G and W refer to GBT and WSRT, respectively. 

\section{Results and Discussion} \label{sec:results}
Here, we discuss our results regarding the column density and corresponding mass, mean line-of-sight velocity, and momentum of the diffuse \hi detected in the CGM pointings in the context of the star formation activities in NGC\,891 and NGC\,4565 (see Table\,\ref{tb:deriv} and Figures \ref{fig:891} and \ref{fig:nhi_v_cgm}). Figures \ref{fig:891} and \ref{fig:nhi_v_cgm} are arranged beginning at the NO\,1W position and proceeding counterclockwise (CCW). The azimuthal angle, $\theta_{\mathrm{CCW}}$, is defined such that $\theta_{\mathrm{CCW}} = 0^\circ$ corresponds to the UP 1J position, with increasing angle measured counterclockwise. The reason for beginning the plotting at NO\,1W is to ease the visualization of the co-rotation of the CGM with the \hi in the disk, in Figure \ref{fig:nhi_v_cgm}. We include the principal axis CGM pointings one GBT beam away from the \hi disk presented in \citetalias{Das2020b,Das2024a}, re-reduced using the techniques described in this paper for completeness, to achieve a full 360\deg picture. 

\subsection{Column density}

\begin{figure*}
\centering
    \includegraphics[width=0.33\textwidth]{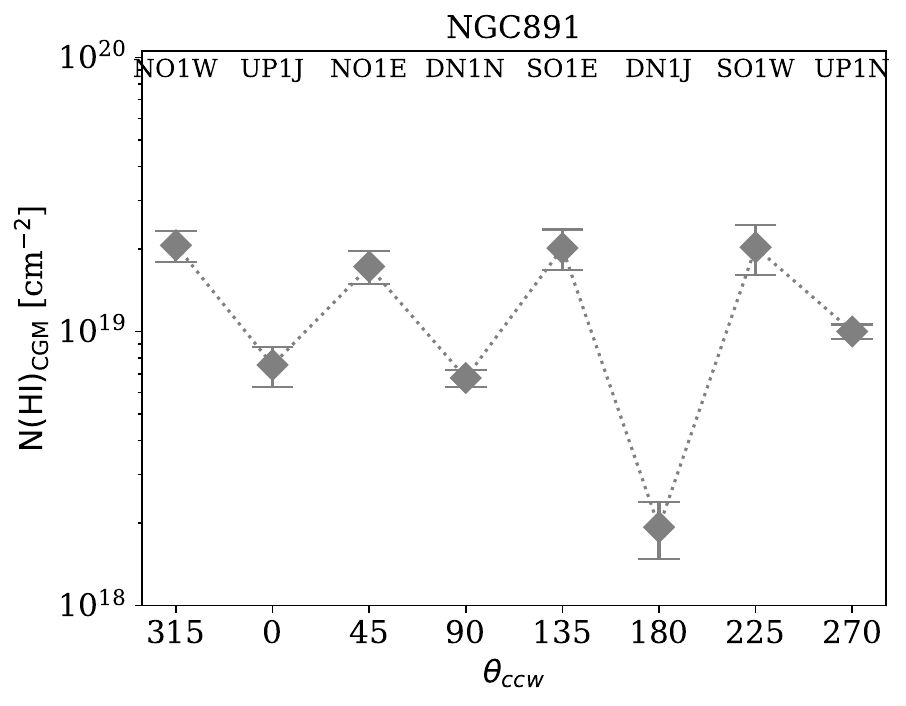}
    \includegraphics[width=0.33\textwidth]{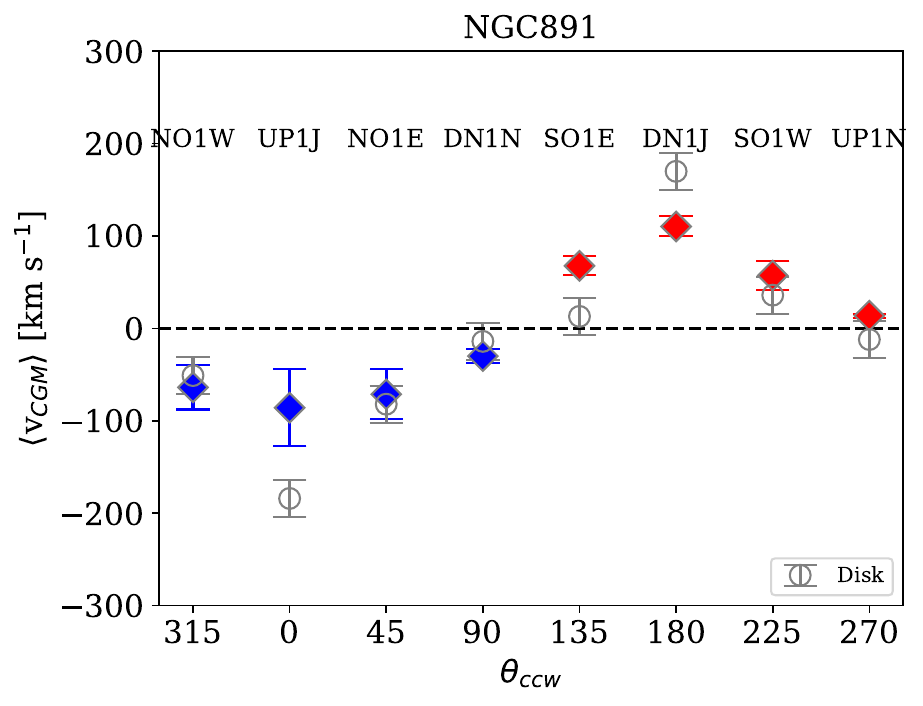}
        \includegraphics[width=0.33\textwidth]{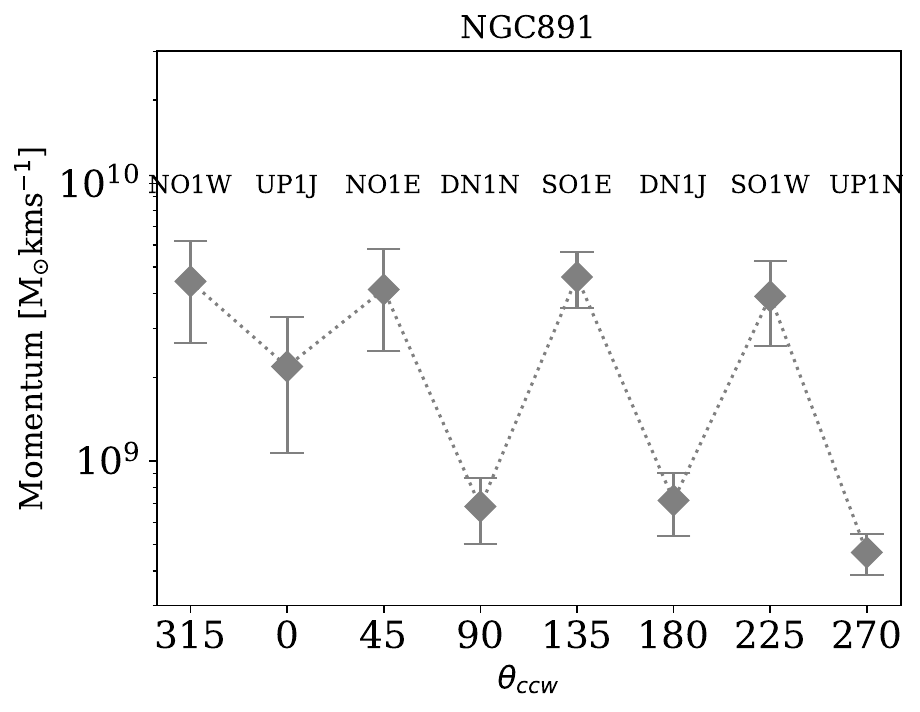}
    \includegraphics[width=0.33\textwidth]{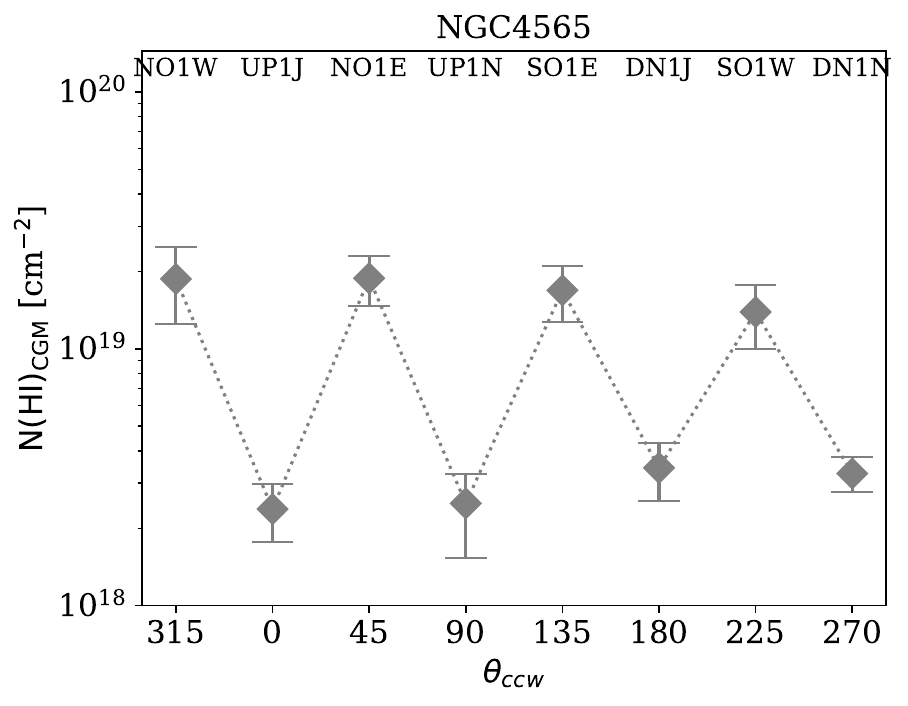}
    \includegraphics[width=0.33\textwidth]{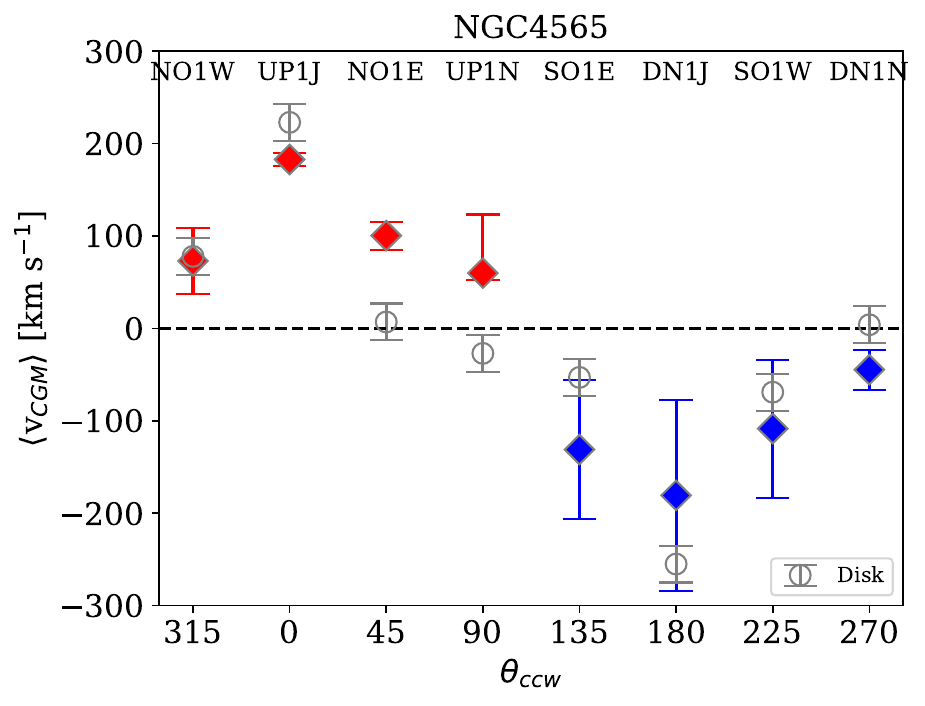}
    \includegraphics[width=0.33\textwidth]{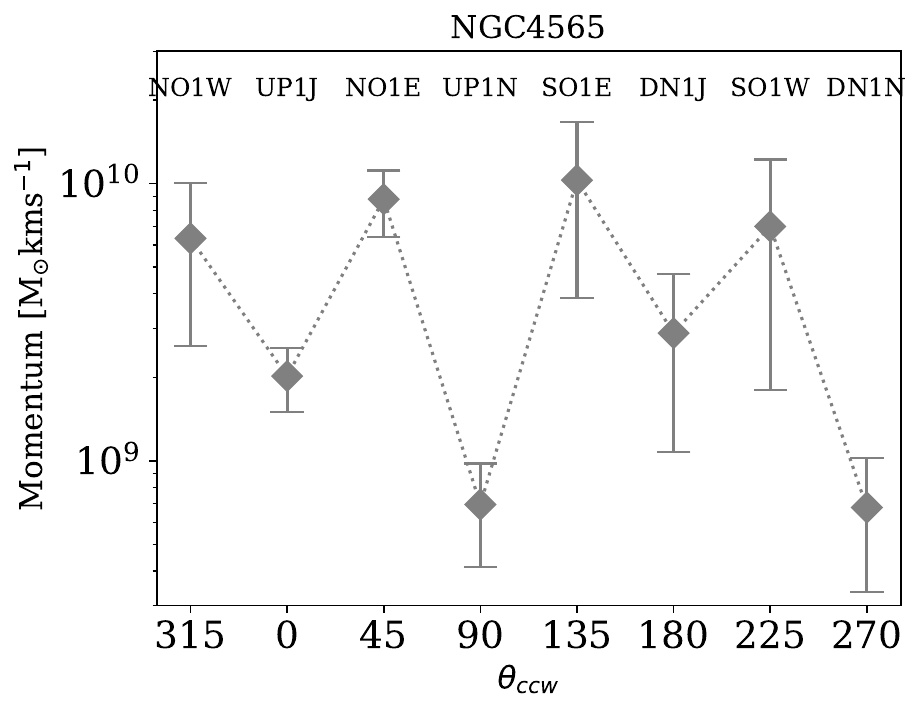}
    \caption{Variation of N(\hin) (left), mean line-of-sight velocities (middle), and mean momentum (right) of the inner CGM of NGC\,891 (top) and NGC\,4565 (bottom) against azimuthal positions, $\theta_{\mathrm{CCW}}$. \textbf{Left:} N(HI) in the CGM along the minor-axis and major-axis pointings is smaller than that along the off-axis pointings, with non-uniformity across galaxies as well as azimuthal positions. However, by design, the minor-axis and off-axis pointings are roughly equidistant from the galaxy, and the major-axis pointings are further away. Thus, there is no correlation of N(HI) in the inner CGM with impact parameter. \textbf{Middle:} Symbols are color-coded to indicate red/blueshifted velocities of circumgalactic \hin. Gray circles correspond to the \hi disk velocity at similar position angles of CGM pointings, v$_{\mathrm{disk}}$. We adopt a standard uncertainty of 20 km\,s$^{-1}$ on v$_{\rm{disk}}$. Note that the velocity profile of the CGM looks qualitatively similar to the disk, but quantitatively, the CGM moves slower/faster than the disk closer to the major/minor axes, revealing a lagged co-rotation. \textbf{Right:} Momentum of the neutral inner CGM is the smallest/largest/intermediate along minor-axis/off-axes/major-axes pointings, showing that its azimuthal variation does not directly correlate to the momentum of the disk, where the major axes are expected to show the maximum momentum.}
    \label{fig:nhi_v_cgm} 
\end{figure*}

For NGC\,891, the global average N(\hin) in GBT and WSRT are $3.39 \pm 0.08 \times 10^{19}$cm$^{-2}$ and $2.08 \pm 0.31 \times 10^{19}$cm$^{-2}$, respectively. And for NGC\,4565, the global average N(\hin) in GBT and WSRT are $3.99 \pm 0.03 \times 10^{19}$cm$^{-2}$ and $2.99 \pm 0.47 \times 10^{19}$cm$^{-2}$, respectively plotted in Figure \ref{fig:891} as green and blue horizontal dashed lines. 
As reported in Table \ref{tb:deriv}, the percentage of emission detected by GBT that cannot be explained by WSRT (GBT$_{\mathrm{\%excess}}$) is $30-38$\% and $20-26$\%  in NGC\,891 and NGC\,4565, respectively, implying the detection of circumgalactic \hi (Figure\,2). The column density of \hi in the CGM is plotted in Figure \ref{fig:nhi_v_cgm} (top panel). 

In NGC\,891 we observe a average N(\hin)$_{\mathrm{CGM}}$ of $1.96 \pm 0.91 \times 10^{19}$cm$^{-2}$, $4.74 \pm 1.88 \times 10^{18}$cm$^{-2}$, $8.38 \pm 1.10 \times 10^{18}$cm$^{-2}$  along the off, major, and minor axes, respectively. In NGC\,4565, we observe an average N(\hin) of  $1.71 \pm 1.33 \times 10^{19}$cm$^{-2}$, $2.91 \pm 1.51 \times 10^{18}$cm$^{-2}$, $2.88 \pm 0.14 \times 10^{18}$cm$^{-2}$  along the off, major, and minor axes, respectively. 

There are two main interesting results from this: 1) off-axis pointings host more N(\hin) compared to the principal axis pointings in both galaxies, and 2) the average N(\hin) in the off-axis pointings is comparable in both galaxies. We break down the possible explanations below by first addressing the possibility of uncharacterized systematics and double-counting emission, then concluding with explanations for a true emission.

\subsubsection{Systematics, or True?}
\label{sec:sysortrue}
Here, we take an agnostic approach to test if the excess \hi detected by the GBT is due to any yet-uncharacterized systematic 
in two ways: different methods of convolving the WSRT-HALOGAS data with the GBT beam and cross-pointing contamination due to sidelobe overlap.

As described in detail in the Appendix \ref{appdx:p18}, we compare spectra from the \citetalias{Pingel2018} GBT data cube. Similar to our observation, the off-axis N(\hin) in \citetalias{Pingel2018} are systematically larger than N(\hin) of the principal axes, driven by stronger \hi disk contamination at off-axis pointings. This rules out the possibility of an unresolved calibration or other systematic in our deep stare GBT observation.

Since the N(\hin)$_{\rm{WSRT}}$ values are convolved with the GBT beam, a possible issue could be an uncharacterized GBT beam systematic. The values reported in Table \ref{tb:deriv} originate from convolving the HALOGAS cubes with a circularized GBT beam and extracting/integrating the spectra at each pointing. We test the effects of different ways to treat the same GBT beam on our results in Appendix\,\ref{appdx:crossinstrument}. We conclude that the circularized GBT beam, along with the systematic uncertainty, is a reasonable representation of the GBT beam.  

It is also possible that the excess emission along off-axes is double-counted (discussed in greater detail in the Appendix \ref{appdx:beam_contam}). 
The positions of the pointings are designed to be at least one GBT FWHM away from the \hi disk edge and other pointings. The N(\hin) of a given CGM pointing may be contaminated by N(\hin) of adjacent pointings through sidelobe overlap, leading to double-counting. We explore this option in Appendix \ref{appdx:beam_contam} and find that the sidelobe overlap is negligible for off-axis pointings.

To the best of our ability, we have thoroughly tested all methods of robustly characterizing the systematics. Based on all the aforementioned methods, we are reasonably confident that this excess \hi emission detected by the GBT compared to the WSRT along the off-axis is physical. 


\subsubsection{Assuming true emission}
\label{trueemission}

\begin{figure}
    \centering
    \includegraphics[width=\linewidth,trim= 2.70 2.50 2.70 2.50,clip]{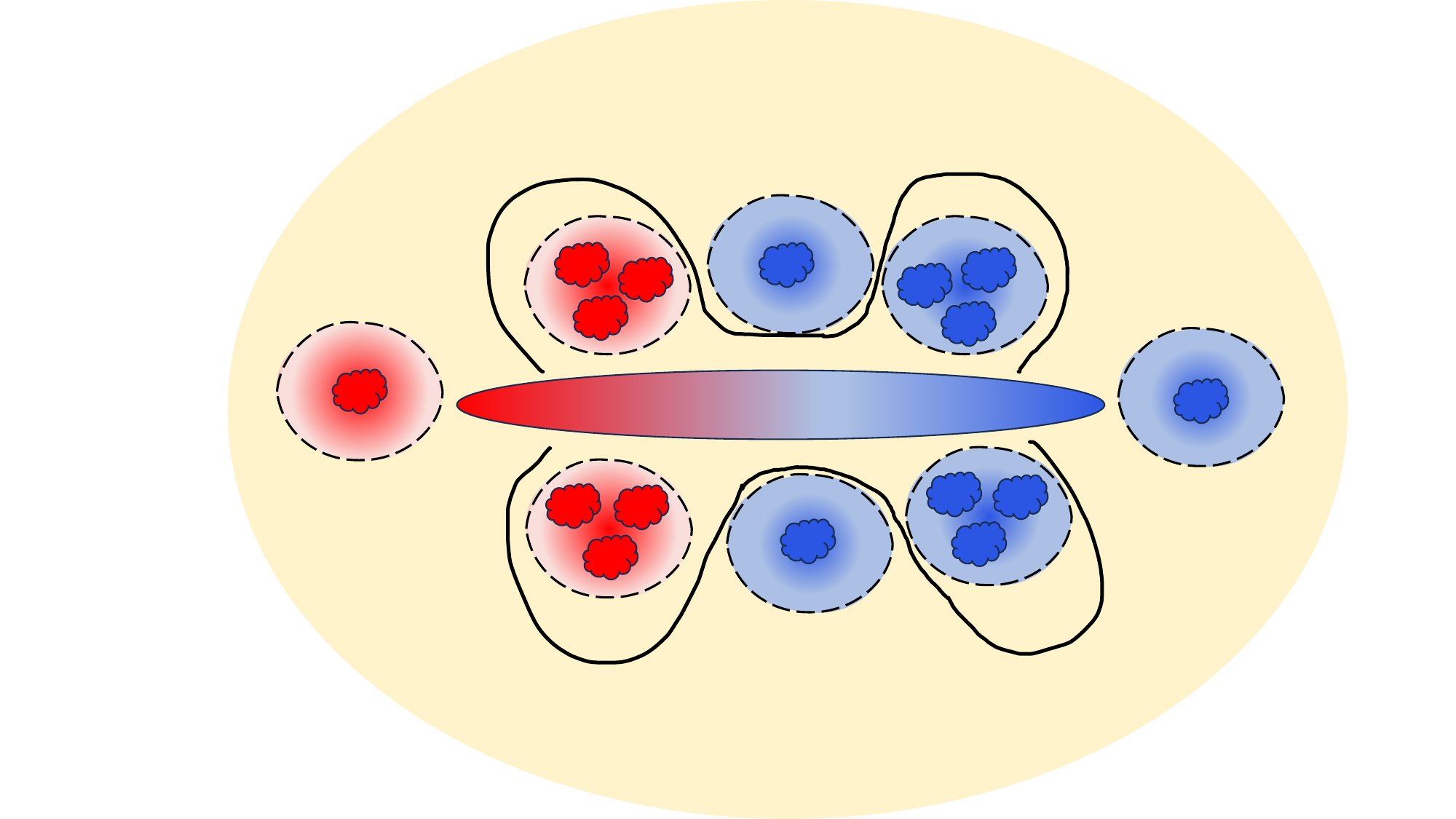}
    \caption{Illustration of the butterfly-like morphology of the N(\hin) of inner CGM (out to $\approx$ 24--30\,kpc from the disk) describing the two possibilities (diffuse beam-filling gas vs clouds smeared by the GBT beam) outlined in Section \ref{trueemission}. The oval represents the \hi disk, and is color-coded red to blue to indicate red/blue-shifted velocities.}
    \label{fig:morph}
\end{figure}

Both galaxies host more \hi in their off-axis compared to their principal axis pointings. This is interesting because this nullifies the assumption of azimuthal symmetry in the circumgalactic \hi distribution, and unveils a significant repository of previously undetected circumgalactic \hin. Assuming the entire detected \hi is beam-filling, the excess \hi along off-axis pointings compared to the principal axis pointings indicates a ``butterfly-like" morphology of \hi in the disk-halo interface (Figure\,\ref{fig:morph}). Part, if not all, of the emission could be smeared out \hi clouds (with each of their N(\hin) below the sensitivity level of HALOGAS), as is often found in high angular resolution follow-ups of diffuse structures \citep[e.g.,][]{BraunThilker2004,Wolfe2013,Wolfe2016}. We emphasize that we cannot concretely claim the size scale of the \hi from these observations alone. High angular resolution observations from interferometers would be required to disentangle the dominant angular scale, or morphology, of the detected \hin. It is important to note that the goal of this and \citetalias{Das2020b,Das2024a} investigations is zeroth-order detections of diffuse gas outside the \hi disk. We consider a few possible explanations for the \hi excess toward off-axis pointings, which gives rise to the ``butterfly morphology" shown in Figure \ref{fig:morph}. 

\begin{enumerate}
    \item \textbf{Ionization:} One explanation for the ``butterfly morphology" is that the degree of ionization varies around the \hi disk with azimuthal angle, leaving different fractions of neutral gas. To explain the excess \hi along off-axis, we consider the possible reasons behind the lack of \hi along the major and minor axes. Along the major axis, \hi is known to truncate in the planar direction but extend farther in the extraplanar region in some galaxies, including NGC\,891 \citep{Oosterloo2007}. Although the origin of the truncation along the major axis is still debated, it is largely assumed that it marks the outer boundary of the gaseous disk. As one steps along the minor axis, and in the case of NGC\,891, strongly ionizes the surrounding gas \citep{Oosterloo2007,Fraternali2008}. This imparts a biconical lack of \hi along the minor axes, which has also been detected in the Milky Way \citep{DickeyLockman1990}. With insufficient stellar or nuclear radiation to ionize the gas, the off-axis remains predominantly neutral, producing the observed \hi excess. 
    \item \textbf{Smeared filaments:} Although ``hot mode" accretion likely dominates in these two galaxies, as they are both massive and lower-$z$, it is not impossible that some of the off-axis \hi is evidence for smeared out filaments from ``cold mode" accretion preferentially along the off-axis \citep{Keres2005}, because, as discussed in the previous paragraph, the gas is ionized along the minor axis. 
    \item \textbf{Kinematic effects:} 
    The observed circumgalactic HI distribution can be described with an analogy of how a rotating lump of clay is shaped into a bowl. Gentle upward pressure from the base (i.e., the \hi disk) draws the clay vertically. Outward pressure from within or inward pressure from outside causes the bowl to flare, generating an azimuthally asymmetric shape. We discuss other possibilities below.
    \begin{enumerate}
        \item \textbf{\textit{Scalloping}}: An increasing scale height of \hi from the center to the edge of the \hi disk due to scalloping (high-amplitude bending waves) is already known to be present in the Milky\,Way \citep{Kulkarni1982}. What we are observing might be an extended low N(\hin) version of that in our target galaxies. 
        \item \textbf{\textit{Warps:}} The diffuse low N(\hin) structure surrounding the HALOGAS-detected high N(\hin) disk could have a significant warping, causing it to be more morphologically similar to a ``butterfly" rather than an extended disk. Precise origins of galactic warps are still an open field of study, but some suggest a tidal disruption \citep{Bailin2003,Weinberg&Blitz2006} or a misalignment of the axis of rotation of the dark matter halo with the disk \citep{Dubinski2009,Han2023}. \hi warps have been detected in many galaxies, including the Milky\,Way \citep{Kerr1958} and in NGC\,4565, which reveals a warped disk with a flaring layer along UP\,1J \citep{Rupen1991, Zschaechner2012}. 
     \item \textbf{\textit{Transitional Boundary:}}     The off-axis may also be a spatial boundary/transition between active and inactive accretion. This excess in off-axis could be a result of a built-up lag in the redistribution of angular momentum. 
    \end{enumerate}
    
\end{enumerate}

We find no significant difference in the total amount of N(\hin) in the two galaxies. Across off-axis pointings, the percentage of off-axis N(\hin) is similar in both galaxies, relative to their principal axis pointings. This is intriguing as it may mean that \hi at the off-axis position is not involved in whatever is causing the discrepancy in star formation in the two galaxies. In \citetalias{Das2024a,Das2020b}, principal axis detections were below the sensitivity limit of HALOGAS. In this case, small clouds could be present but undetectable by HALOGAS. Since the off-axis detections are above the sensitivity of HALOGAS ($\sim$ 10$^{19}$), this implies the presence of extended CGM structure.

\subsection{Mass}
We discuss the results for the mass at each CGM pointing, which is calculated as described in \S \ref{sec:datared}. The explanation for the excess mass and N(\hin) is similar, as they are directly proportional quantities.  We break the results up into four ways for off-axis: NO, SO, E, W, and total mass. Then we compare with principal axis pointings, and then we compare with the total amount seen in the disk for both galaxies.

\subsubsection{NGC\,891}
In the NO, we observe $12.77 \pm 1.20 \times 10^7$\msun, and in the SO we observe $13.63 \pm 1.81 \times 10^7$\msun. There is no significant difference in the M(\hin)$\rm_{CGM}$ of the NO or SO. In the W, we observe $13.79 \pm 1.67 \times 10^7$\msun, and in the E we observe $12.61 \pm 1.40 \times 10^7$\msun. Similarly to the NO/SO, there is no significant difference in the M(\hin)$\rm_{CGM}$ of the W or E. In fact, the \hi mass measured at all off-axis pointings is within 1$\sigma$ of each other. In total, we observe $26.40 \pm 2.18 \times 10^7$\msun in the off-axis. 

Comparing to the previous results of \citetalias{Das2024a}, the M(\hin)$\rm_{CGM}$ along and around the major and minor axes are $3.20 \pm 0.45 \times 10^7$\msun~and $5.65 \pm 0.26 \times 10^7$\msun, resulting in a total mass of $8.85 \pm 0.52 \times 10^7$\msun. 

There is the most mass in the off-axis pointings, followed by the minor and major axes. That means we underestimated the total mass by a factor of 2 to 3 by assuming the off-axis mass can be linearly interpolated from the major and minor axes in \citetalias{Das2024a}. The updated total mass of the inner CGM, including principal axis and off-axis mass measurements, is $35.25 \pm 2.24 \times 10^7$\msun.

The mass of the disk is $4.1 \times 10^9$\msun \citep{Heald2012}. Therefore, the inner CGM is 8.6 $\pm 0.5\%$ of the disk. Of this percent, 6.4 $\pm 0.5\%$ is contributed by the off-axis.

Using the SFR from \cite{Heald2012}, 2.2 M$_\odot$yr$^{-1}$, the updated depletion time scale, $\rm \tau_{dep} =$ (M(\hin)$\rm_{disk}$ + M(\hin)$\rm_{CGM})/SFR$, is 2.02 $\pm 0.01$ Gyr, consistent with the previously found $2.0\pm0.8$\,Gyr in \citetalias{Das2024a}. Although the $\rm \tau_{dep}$ calculated in this work is consistent with that from \citetalias{Das2024a}, we note that the latter $\rm \tau_{dep}$ from \citetalias{Das2024a} also includes pointings greater than one FWHM GBT beam away from the disk (i.e., DN\,2J, DN\,3J, etc). The $\rm \tau_{dep}$ implies that, unless there is another repository of \hi gas, NGC\,891 will deplete its \hi fuel reserves in 2 Gyr, ceasing/lessening star formation, or the rest of the accreted \hi is ionized, especially along the principal axis (see discussion in Section \ref{sec:results}. \citetalias{Das2024a} suggests the possibility of in-situ fueling that may explain some of the lack of fuel available to form stars.

\subsubsection{NGC\,4565}
In the NO and SO, we find masses of $17.426 \pm 3.48 \times 10^7$\msun and  $14.29 \pm 2.64 \times 10^7$\msun. In the W and E, we observe $15.13 \pm 3.41 \times 10^7$\msun and $16.57 \pm 2.74 \times 10^7$\msun. This yields a total mass of $31.71 \pm 4.38 \times 10^7$\msun along the off-axis. Similarly to NGC\,891, we find no significant deviation in the mass of the NO, SO, W, and E pointings. 

Comparing to the principal axis pointings, M(\hin)$\rm_{CGM}$ along and around the major and minor axes (out to around 1 GBT beam pointing away) are $2.70 \pm 0.50 \times 10^7$\msun~and $2.68 \pm 0.51 \times 10^7$\msun, resulting in a total mass of $5.38 \pm 0.71 \times 10^7$\msun. 

In NGC\,4565, we see a similar trend as in NGC\,891; the off-axis pointings have a greater N(\hin) than the major or minor axis pointings. Considering the mass from the off-axis and all principal axis pointings, the total mass is $37.46 \pm 4.44 \times 10^7$\msun. 

The mass of the disk is $7.3 \times 10^9$\msun \citep{Heald2012}. The M(\hin)$\rm_{CGM}$ is about 5.1 $\pm 0.6\%$ that of the disk. The off-axis pointings contribute 4.3 $\pm 0.6\%$ of that mass. The SFR for NGC\,4565 is 0.67 M$_\odot$yr$^{-1}$, from \cite{Heald2012}. The updated depletion time scale, 
$\rm \tau_{dep} =$ (M(\hin)$\rm_{disk}$ + M(\hin)$\rm_{CGM})/SFR$ is 11.45 $\pm$ 0.07 Gyr. This implies that NGC\,4565 has plenty of \hi mass to accrete and form stars over the next 11 Gyrs.

\subsubsection{Both galaxies}
In both galaxies, we detect 1) low column density gas (larger by 2--3 times than \citetalias{Pingel2018}) at all CGM pointings, 2) 2-3x more \hi along the off-axis compared to the principal axis pointings, and 3) no significant deviation in the mass of \hi found at each off-axis pointing. When comparing the two galaxies, the \hi along the off-axis is similar (within 1$\sigma$) in both galaxies. This is intriguing because previously \citetalias{Das2020b, Das2024a} found that the major axis hosted more \hi compared to the minor axis, and NGC\,891 hosts overall more \hi than NGC\,4565 (this is obvious in the top panel of Figure \ref{fig:nhi_v_cgm}). However, our findings show that the off-axis hosts more mass than the major axes, but hosts similar mass across the two galaxies. This implies the \hi along the off-axis may be a ubiquitous trait among spiral galaxies. 

\subsection{Velocity}
In Figure \ref{fig:nhi_v_cgm}, we plot the mean line-of-sight velocities of each pointing. We notice three interesting trends in both galaxies: 1) presence of an inflated diffuse co-rotating \hi disk, 2) difference in v$_{\rm{CGM}}$ scale compared to the v$_{\rm{disk}}$, and 3) differences in the velocities of CGM pointings.

The velocity profile of the \hi in both galaxies, which here extends out to 20 to 30 kpc from the disk, for NGC\,891 and NGC\,4565 respectively, is reminiscent of the shape of their already well-characterized \hi disk velocity profiles (gray points on Figure \ref{fig:nhi_v_cgm}). The major axis pointings achieve the largest velocity. The minor axis pointings are closest to zero because of a lack of angular momentum transfer. The off-axis pointing is between these two trends. Since we are 20 to 30 kpc from the disk, these velocity profiles are consistent with an extended diffuse co-rotating \hi disk.

As noticed in \citetalias{Das2020b} and \citetalias{Das2024a}, although the overall shape of the velocity profile is consistent with the disk, the ``scale" of the v$_{\rm{CGM}}$ as compared to v$_{\rm{disk}}$ is different across both galaxies and each pointing. NGC\,891 v$_{\rm{CGM}}$ values are generally either roughly consistent ($\sim$ 1 $\sigma$) with the v$_{\mathrm{disk}}$ or less (described previously in \citetalias{Das2024a} as tighter), except in the case of SO\,1E. This makes sense both because there is a known, well-characterized velocity lag in \hi along the minor axis of NGC\,891 \citep{Fraternali2008}. The lag in velocity as a function of scale height can be caused by internal mechanisms and/or by angular momentum conservation. 

Despite \cite{Zschaechner2012} also showing a lag, NGC\,4565, shows somewhat of the opposite trend in that the v$_{\mathrm{CGM}}$ are generally consistent with or greater than v$_{\mathrm{disk}}$, except UP\,1J. Although, as previously noted, NGC\,4565 has a warp and flaring layer along UP\,1J, which may contribute to the higher observed velocity along UP\,1J. Velocities greater than v$_{\mathrm{disk}}$ are intriguing and may have something to do with the low-level AGN present in NGC\,4565. The AGN in NGC\,4565 may be injecting energy into the CGM (or possesses an outflow) along the minor-axis pointings, and subsequently has not had enough time (and /or there is an active faucet of energy) for the minor-axis pointings to fully distribute total momentum to the major axis, where it would be accreted to form stars. On the other side, there may be some internal mechanisms, in situ fueling or the lack of an AGN, that have allowed NGC\,891 to transfer momentum more efficiently, thus allowing for lower velocities and therefore more opportunities for cooler gas to be channeled to the disk to later form stars.

In the previous paragraph, we summarized broad trends in velocity but noted exceptions in both galaxies. These exceptions embody the importance of this work, which is the lack of azimuthal symmetry. Likely, each pointing has unique explanations. Explanations for velocities that are not consistent with v$_{\mathrm{disk}}$ could be internal mechanisms, like outflow, fountains, inflows, or external processes like low-level ram pressure stripping, or the presence of satellite galaxies. 

\subsection{Momentum}
In the bottom panel of Figure \ref{fig:nhi_v_cgm}, we plot line-of-sight momentum against angular position about the galaxy. We notice two trends: 1) off-axis in both galaxies, an overall higher total momentum, and 2) there is a lower spread in NGC\,891 as compared to NGC\,4565. 

The average momentum across the off, major, and minor axes is 4.3 $\pm 0.7\times 10^9$\msun km$s^{-1}$, 1.4 $\pm 0.6\times 10^9$\msun km$s^{-1}$, 0.6 $\pm 0.1\times 10^9$\msun km\,s$^{-1}$ for NGC\,891. The average momentum across the off, major, and minor axes is  8.1 $\pm 2.3\times 10^9$\msun km$s^{-1}$, 2.4 $\pm 0.9\times 10^9$\msun km$s^{-1}$, 0.7 $\pm 0.2\times 10^9$\msun km$s^{-1}$  for NGC\,4565. The greater momentum in the off-axis is primarily driven by the \hi mass. NGC\,891 has a spread (maximum - minimum) of about 1.6x the mean value, while NGC\,4565 exhibits a larger spread of about 2x the mean. We suspect this could be a consequence of NGC\,4565 having a LINER AGN. NGC\,4565 has not yet had ample time to redistribute the momentum from the minor axes to the disk. 



\section{Conclusions}
In this paper, we extend and improve the study of \citetalias{Das2020b,Das2024a} to present a 360\deg view of the inner CGM ($\sim$20-30 kpc) of two nearby edge-on spiral galaxies, NGC\,891 and NGC\,4565. We use the GBT to observe the off-axis region, 45\deg between minor and major axis pointings at one GBT FWHM beam away from the \hi disk using the deep stare technique, observing 3-4 hours per pointing, achieving a 1$\sigma$ sensitivity of 10$^{16}$ \cmsq over a 20 km$^{-1}$ line width. 
By using the techniques described in \citetalias{Pingel2018,Das2020b,Das2024a}, we compare our GBT observations with interferometric WSRT maps from the HALOGAS survey, and investigate the amount of diffuse circumgalactic gas present at the off-axis pointing of NGC\,891 and NGC\,4565. We discuss at length in the appendix different approaches to the systematic uncertainty calculation in this study. 

We describe our science results below:
\begin{enumerate}
    \item $30-38$\% ($20-26$\%) of the emission detected by the GBT at off-axis pointings cannot be explained by WSRT for NGC\,891 (NGC\,4565), showing the detection of circumgalactic diffuse \hin. 
    \item Both galaxies host more N(\hin) in the off-axis pointings as compared to the principal axis pointings, implying that the distribution of \hi is not azimuthally symmetric. This unveils a significant missed \hi repository, previously underestimating the total mass of the \hi in the inner CGM by a factor of $\approx$ two in both galaxies.
    \item Together with the principal axes pointings \citepalias{Das2020b,Das2024a}, our observation reveals the presence of a diffuse ($\lesssim 10^{19}$cm$^{-2}$) inner CGM (20-30\,kpc) co-rotating with the \hi disk in both galaxies.
\end{enumerate}

Despite the high time cost, deep stare observations provide us with a neat spatial distribution of the diffuse circumgalactic \hi around individual galaxies. A deep map would allow us to disentangle whether the emission is a result of smoothed-out filamentary/cloud structures or truly diffuse gas that is warped. This would have significant consequences on the theory of spiral galaxy evolution, dark matter halos, and star formation as a whole. As well, to deeply probe the off-axis of other types of galaxies, perhaps in a greater range of masses and star formation rates, to see if we find ubiquitous trends. 

\section*{acknowledgments}
M.R. acknowledges support from the Deans' Fellowship from the University of Notre Dame Graduate School. S.D. is grateful for the support provided by NASA through the NASA Hubble Fellowship grant HST-HF2-51551.001-A awarded by the Space Telescope Science Institute, which is operated by the Association of Universities for Research in Astronomy, Incorporated, under NASA contract NAS5-26555. M.R. thanks Tom Oosterloo and Nissim Kanekar for helpful discussions at the Hungry HIppos: HI as a Tracer of Galactic Accretion, Interactions and Feedback conference, attending which was made possible by the AAS International Travel Grant and the Zahm Graduate Student Professional Development Award at the University of Notre Dame. M.R. also thanks Nicolas Lehner, J. Christopher Howk, and Saloni Deepak for helpful discussions that improved the manuscript.

The Green Bank Observatory is a facility of the National Science Foundation operated under a cooperative agreement by Associated Universities, Inc. This research made use of data from the Westerbork Synthesis Radio Telescope (WSRT) HALOGAS-DR1. The WSRT is operated by ASTRON (Netherlands Institute for Radio Astronomy) with support from the Netherlands Foundation for Scientific Research NWO. This research has made use of NASA's Astrophysics Data System Bibliographic Services.

\section{Data Availability}
HALOGAS data cubes are open source on the  \href{https://www.astron.nl/halogas/index.php}{HALOGAS website}. Raw GBT spectra can be accessed in the NRAO archive. The files to reproduce the results from this work can be found in Zenodo.

\facilities{GBT, WSRT}

\software{APLpy \citep{Robitaille2012}, AstroPy \citep{Astropy2022}, GBTIDL \citep{Marganian2006}, Matplotlib \citep{Hunter2007}, NumPy \citep{numpy2020}, pybaselines \citep{pybaselines}, \href{https://github.com/astropy/pyregion}{pyregion}, python math \citep{mathpy2020}, SciPy \citep{scipy2022}, Spectral cube \citep{Robitaille2016}}

\appendix
\counterwithin{figure}{section}

\section{Stitching baseline in HALOGAS cubes}
\label{appdx:stitchbaseline}
After convolving the masked HALOGAS cubes with the circularized GBT beam, we extract the spectrum at the coordinates of each off-axis pointing. When analyzing the GBT spectra, we use an iterative baseline fitting procedure described in \citetalias{Das2024a}, which establishes the optimal velocity range of the baseline, $\Delta$v$\mathrm{_{base}}$, and the velocity range of the detected emission, $\Delta$v$\mathrm{_{emit}}$,  (see Table\,1). For a one-on-one comparison between the GBT and HALOGAS data, the spectra from both instruments need to have the same velocity range. However, in NGC\,4565, the velocity baseline in the extracted HALOGAS data is significantly shorter than the baseline in our GBT data; see the top panel of Figure \ref{fig:appdx_velosig}. To remedy this, we stitch a baseline of zeros (reflecting the values of the existing baseline in the masked WSRT spectra) to match the $\Delta$v$\mathrm{_{base}}$ from GBT. Although the short baseline most impacted NGC\,4565, for completeness, we follow the same approach in NGC\,891.

\section{Velocity edge corrections} 
\label{appdx:velocorrect}
As discussed later in Appendix \ref{appdx:Tbscaling}, we implement a channel-wise brightness-temperature scaling between the HALOGAS and GBT data. For that, it is pertinent that the velocity channels in GBT and WSRT data align. 

As mentioned in \S\ref{sec:comp_interfer}, we subtract the systemic velocity of each galaxy taken from \cite{Heald2011} to obtain the WSRT/HALOGAS spectra at the rest frame of the galaxy and reconstruct them at the same velocity channels of the GBT spectra through interpolation. As a result of co-rotation with the disk, both spectra exhibit two distinct edges; we refer to the ``rising" and ``falling" edges interchangeably as the left (negative) and right (positive) edges of the spectra, respectively (top panel of Figure \ref{fig:appdx_velosig}). The rising and falling edges of the \hi emission of the WSRT data in the central pointings do not line up with the GBT data, likely due to a mismatch in the systemic velocities. We find that the velocity alignment of the central pointing edges differs significantly. We treat them separately as described below, but in a two-process approach (velocity width alignment and velocity offset correction).

Line-of-sight velocity is defined from wavelength ($\lambda$) or frequency ($\nu$) Doppler shifts, following two conventions, optical and radio. At the redshifts of NGC\,891 and NGC\,4565, the difference between the optical and radio conventions is less than 1 km/s, smaller than the velocity resolution of both WSRT and GBT data. As a result, we conclude that different conventions in velocity are not the dominant cause for the mismatch between GBT and HALOGAS spectra.



We first check if the velocity width of the GBT and WSRT spectra is within 2 channel widths, corresponding to 5 and 10  km\,s$^{-1}$ for off- and principal axes pointings, of each other. Widths mismatched beyond this threshold cannot be truly aligned, as the alignment of the rising edge will result in the misalignment of the falling edge. If the velocity widths do not deviate by more than 2 channel widths, we apply a velocity edge alignment (middle panel of Figure \ref{fig:appdx_velosig}). We align the falling edge (right side) by finding the velocity offset correction needed to align GBT and WSRT spectra at 25\% of the maximum T$_{\mathrm{B}}$ edge value. The choice of 25\% threshold as the best alignment is decided through trial and error.  The nominal velocity resolution of HALOGAS is 4.12 km\,s$^{-1}$ for NGC 4565 and 8.24 km\,s$^{-1}$ for NGC 891. The mismatch in velocity widths from GBT and HALOGAS is likely due to differences in the velocity resolution. Although NGC 891 does not align in velocity width for off-axis pointings based on our 2 velocity channel metric, the difference is less than the velocity resolution of HALOGAS.

Because the velocity widths of GBT and HALOGAS emission are consistent with each other in the central pointing of NGC\,4565, we apply a velocity offset correction factor to align the rising and falling edges of HALOGAS spectra with the GBT spectra. On the other hand, the velocity widths in the central pointing of NGC\,891 in GBT and HALOGAS are not consistent (the difference is $>$ 2 channel widths, 5 km\,s$^{-1}$). Thus, no velocity offset adjustment would properly align both edges simultaneously. Therefore, we do not apply a velocity offset correction factor to the central pointing of NGC\,891. Different outcomes in the two galaxies imply that the mismatch between our GBT spectra and WSRT spectra convolved with the GBT beam is not due to any uncertainty in the GBT beam response, because that would have affected both galaxies the same way.

The trend of misaligned velocities percolates through to the off-axis pointings. NGC\,4565 requires a velocity offset correction of the equivalent magnitude and sign that was needed for the central pointing. We only apply this correction as needed. NGC\,891 did not need any additional velocity correction in the off-axis CGM pointings. These corrections are enumerated in Table \ref{tb:deriv}, as v$\rm_{offset}$, and visually represented in the middle panel of Figure \ref{fig:appdx_velosig}. 

\section{Channel-wise Brightness Temperature Scaling}
\label{appdx:Tbscaling}
To compare the CGM measurements from HALOGAS and GBT data, we need to ensure that the measurements of the \hi disk are similar from the two instruments. However, HALOGAS observations of the central pointing are larger than those from the GBT. To account for this mismatch, we apply a brightness temperature scaling factor, $f_{\rm{scalar}}$. \citetalias{Das2020b,Das2024a} obtained $f_{\rm{scalar}}$ from the ratio of integrated intensity of the \hi disk from HALOGAS and GBT. It works under the assumption that $f_{\rm{scalar}}$ is the same across all velocity channels, resulting in an overall loss of information about channel-wise differences. In this work, we move one step further and determine the scaling factor between HALOGAS and GBT in each velocity channel, $f_{\rm{channel}}$, thereby preserving the channel-wise differences and enhancing robustness.

\begin{figure}
\includegraphics[width=1.0\columnwidth]{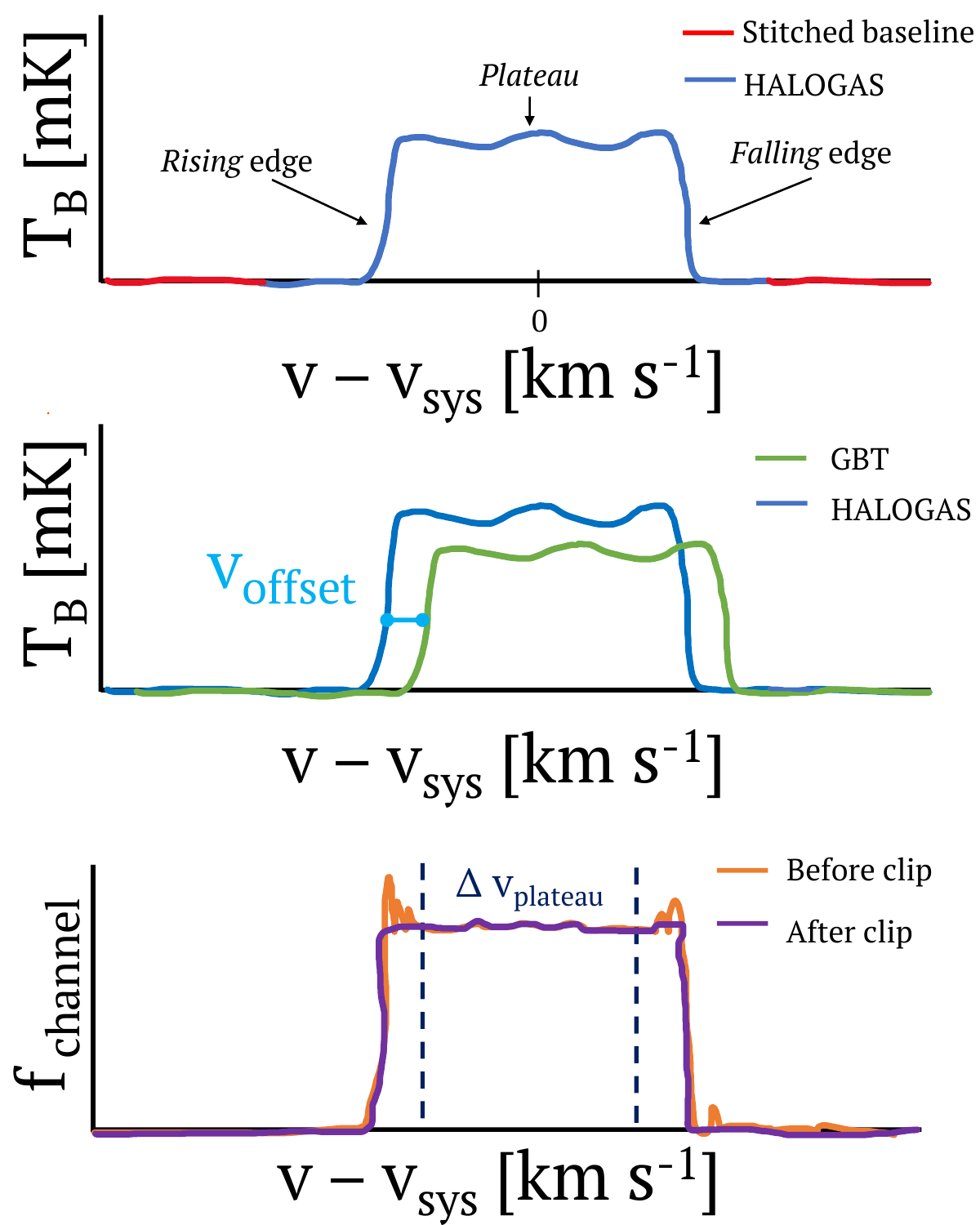}
\caption{An illustration of baseline stitching (Appendix \ref{appdx:stitchbaseline}, top panel), velocity offset corrections (Appendix \ref{appdx:velocorrect}, middle panel), and sigma clipping (Appendix \ref{appdx:Tbscaling}, bottom panel).}
\label{fig:appdx_velosig}
\end{figure}

We calculate the channel-wise brightness temperature scaling factor, $f_{\rm{channel}}$, by dividing the central GBT spectrum by the central WSRT spectrum. We plot an illustrated version of $f_{\rm{channel}}$ in the bottom panel of Figure \ref{fig:appdx_velosig}. We apply $f_{\rm{channel}}$ to all observed HALOGAS CGM pointings to ensure they are properly scaled to their central pointing. When plotting $f_{\rm{channel}}$ as a function of velocity, we notice some channels with spurious values; see ``Before clip" in the bottom panel of Figure \ref{fig:appdx_velosig}. The presence of these spurious values is a consequence of imperfect velocity edge corrections (see Appendix\,B). Any misalignment in the velocity edges, resulting mainly from the different velocity widths in the convolved WSRT spectra and the GBT spectra, can lead to values significantly smaller or larger than one close to the velocity edges. The final WSRT spectrum to be compared with our GBT spectrum is the channel-wise multiplication of $f_{\rm{channel}}$ (from the central pointing) to the GBT beam-convolved WSRT spectra. Thus, the spurious value is further carried through and contributes non-negligibly when integrated across channels to determine the N(\hin), resulting in an unphysical excess or deficit in \hi emission. 

The location and extent of the spurious values cover a large range. Finding a common way to mitigate most spurious values at all pointings in both galaxies proved to be extremely challenging. Therefore, we decided to apply a two-step clipping method. One can decompose the $f_{\rm{channel}}$ spectrum into three regions (top panel of \ref{fig:appdx_velosig}): 1) the flat \textit{baseline}, 2) the $\sim$linear rising and falling velocity \textit{edges}, and 3) the \textit{plateau} between the rising and falling \textit{edges} featuring approximately the mean scaling factor. In the first step, we remove large outlier channels on the rising and falling \textit{edges} using difference sigma clipping. In the second step, we use a global sigma clipping to smooth out any missed channels with spurious values, especially those in the \textit{baseline} and the \textit{plateau}.

\textbf{Difference Sigma Clip:} We calculate the nearest-neighbor channel-wise difference in brightness temperature, $\delta f_{\rm{channel}}$. We calculate the mean and standard deviation of $\delta f_{\rm{channel}}$. We then identify the spurious channels corresponding to a $\pm 3\sigma$ difference, then remove and replace them with cubic interpolation. We tested different interpolation functions: linear, quadratic, cubic, and nearest-neighbor, and found that the cubic interpolation best preserves the original shape of $f_{\rm{channel}}$. 

\textbf{Global Sigma Clip:} After removing the outliers in rising and falling edges, we apply a global sigma clip. To establish the upper and lower bounds of the sigma clip, we determine a set of velocities corresponding to an ``uncontaminated" region, $\rm \Delta v_{plateau}$ in the bottom panel of Figure \ref{fig:appdx_velosig}, within which we can calculate the mean and standard deviation, $\mu$ and $\sigma$. We determine this velocity range, $\rm \Delta v_{plateau}$, iteratively, first with a small range centered about 0, -50 to 50. We move outwards in steps of of 10 kms$^{-1}$, in each instance calculating $\mu$ and $\sigma$ of $f_{\rm{channel}}$ (beginning after 5 initial steps). The iterative velocity step forward stops when $\mu$/$\sigma$ exceeds 0.1, i.e., 10\%. The convergent range for all pointings in both galaxies turns out to be $\rm \Delta v_{plateau} =$ [-200,200] kms$^{-1}$. We calculate the final $\mu$ and $\sigma$ of $f_{\rm{channel}}$ over this $\rm \Delta v_{plateau}$ free of channels with spurious values. Then, we step through the entire spectrum, and adopting the \href{https://numpy.org/doc/stable/reference/generated/numpy.clip.html}{\texttt{numpy.clip}}
method, reject all values below 0.001 and more than 3 $\sigma$ away from $\mu$.  The final result of the combined difference and global sigma-clipped is represented as ``After clip" in Figure \ref{fig:appdx_velosig}. This method allows us to retain the information of the channel-wise scaling between GBT and WSRT, but mitigates spurious channels resulting from residual velocity misalignment between the two instruments. We tested these new steps on the principal axes pointings presented in \citetalias{Das2020b,Das2024a} and found negligible differences.

\setlength{\fboxsep}{4pt}  
\setlength{\fboxrule}{0.8pt} 


\section{Testing different methods of cross-instrument comparison}
\label{appdx:crossinstrument}

\begin{figure*}
\centering
\fbox{%
  \scalebox{0.8}{%
    \begin{minipage}{0.9\linewidth}
        \centering
        \includegraphics[width=\linewidth, trim=0 0.50in 0 0, clip]{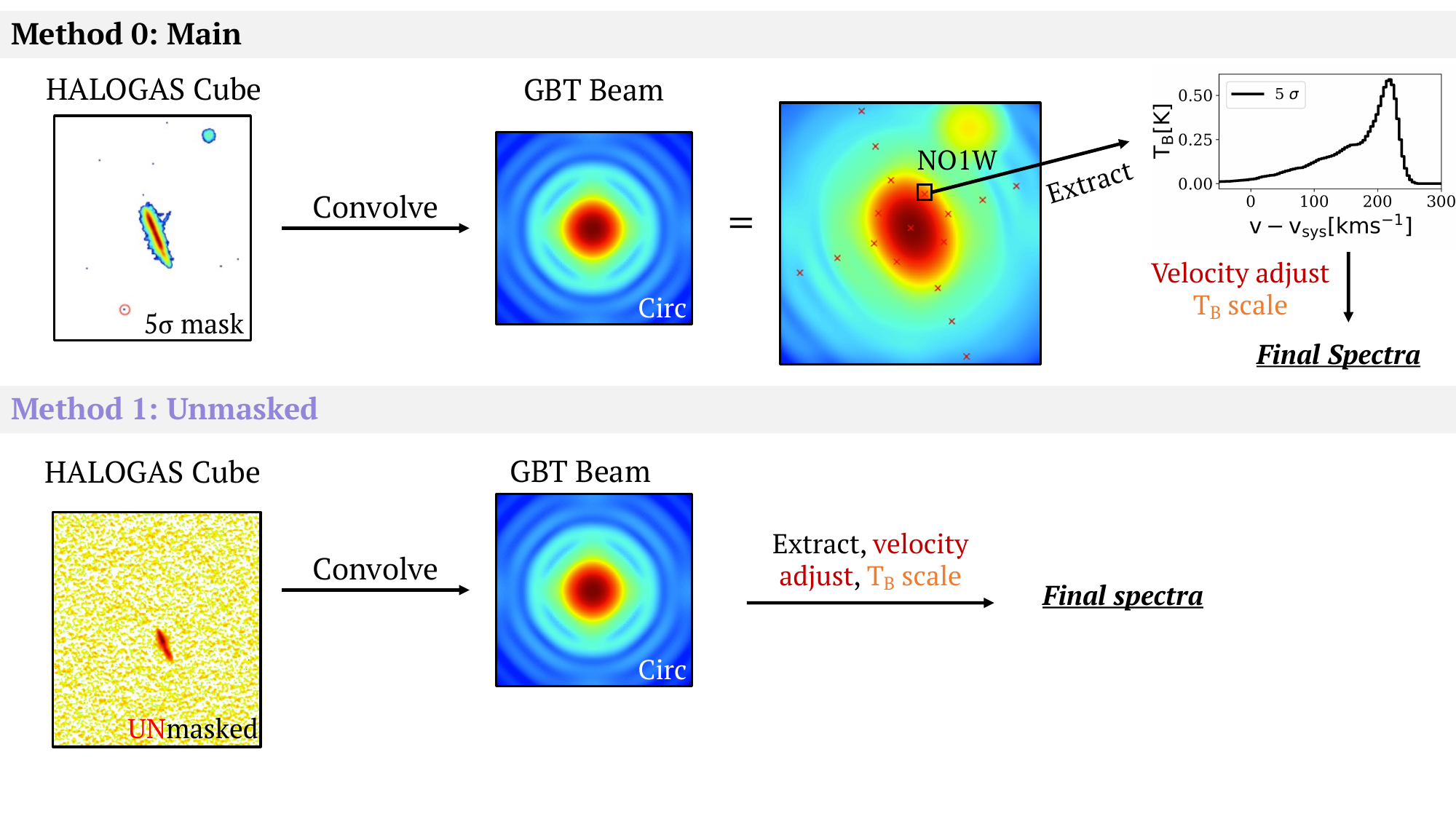}

        \includegraphics[width=\linewidth, trim=0 1.75in 0 0, clip]{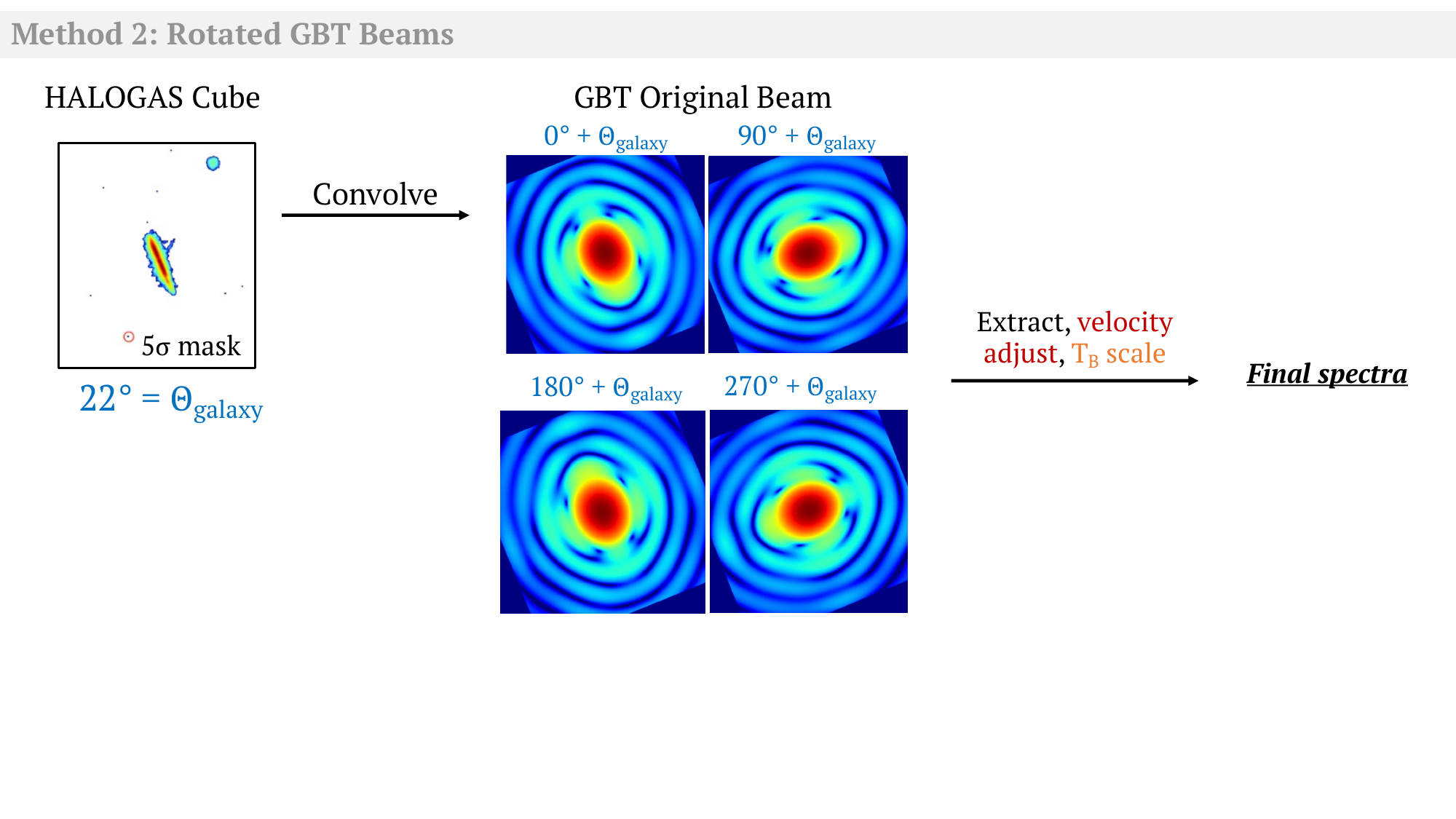}

        \includegraphics[width=\linewidth, trim=0 2.5in 0 0, clip]{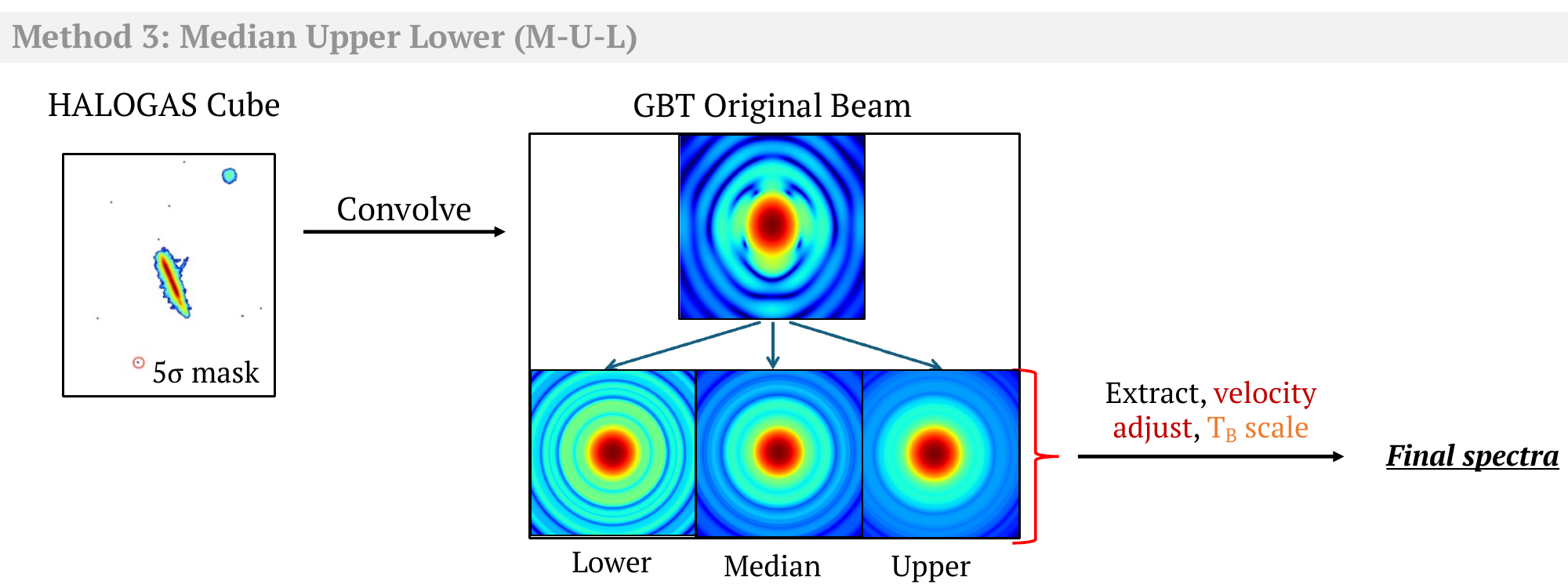}
    \end{minipage}
  }%
}
\caption{Illustration of different (main and alternate) methods convolving HALOGAS cubes with GBT beam maps. The results and interpretations of all methods are described in Appendix \ref{appdx:crossinstrument}.}
\label{fig:appdx_allmeth}
\end{figure*}

\begin{figure*}
\centering
\includegraphics[width=0.45\linewidth]{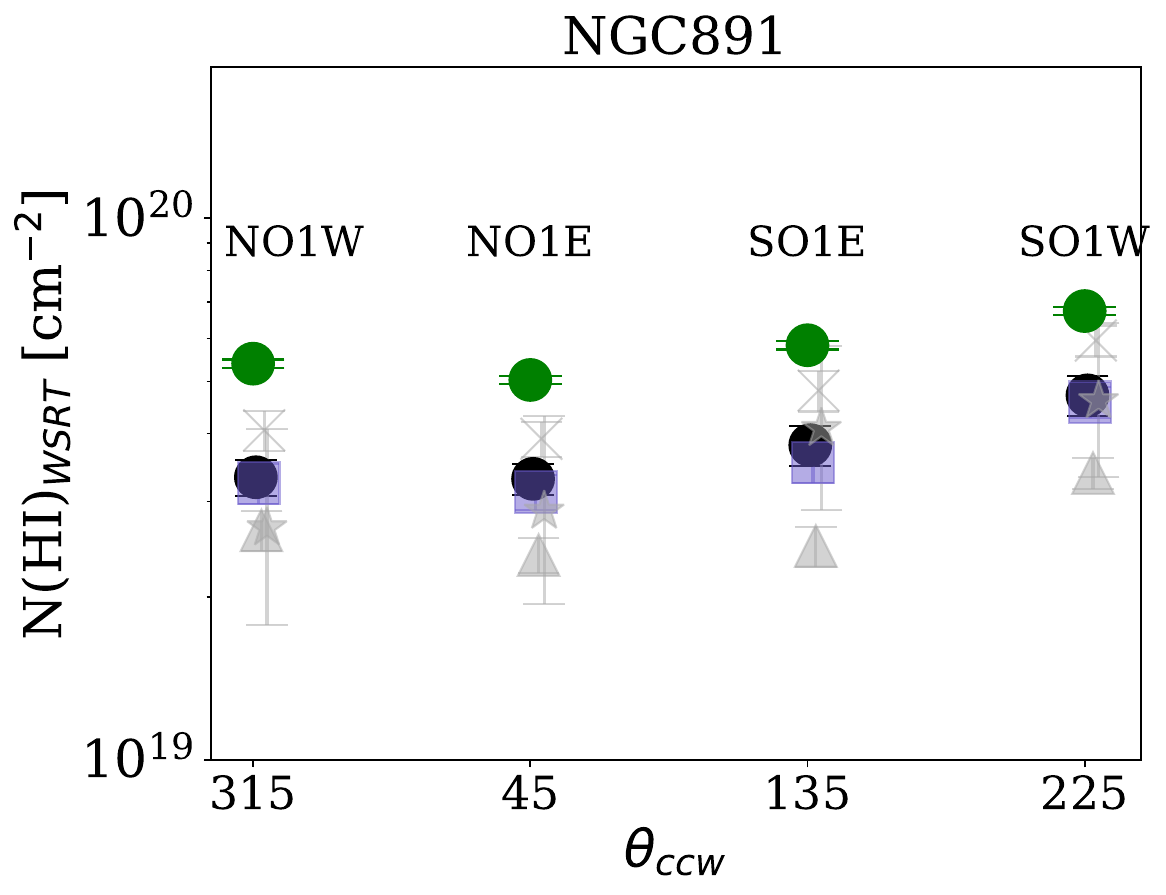}
\includegraphics[width=0.45\linewidth]{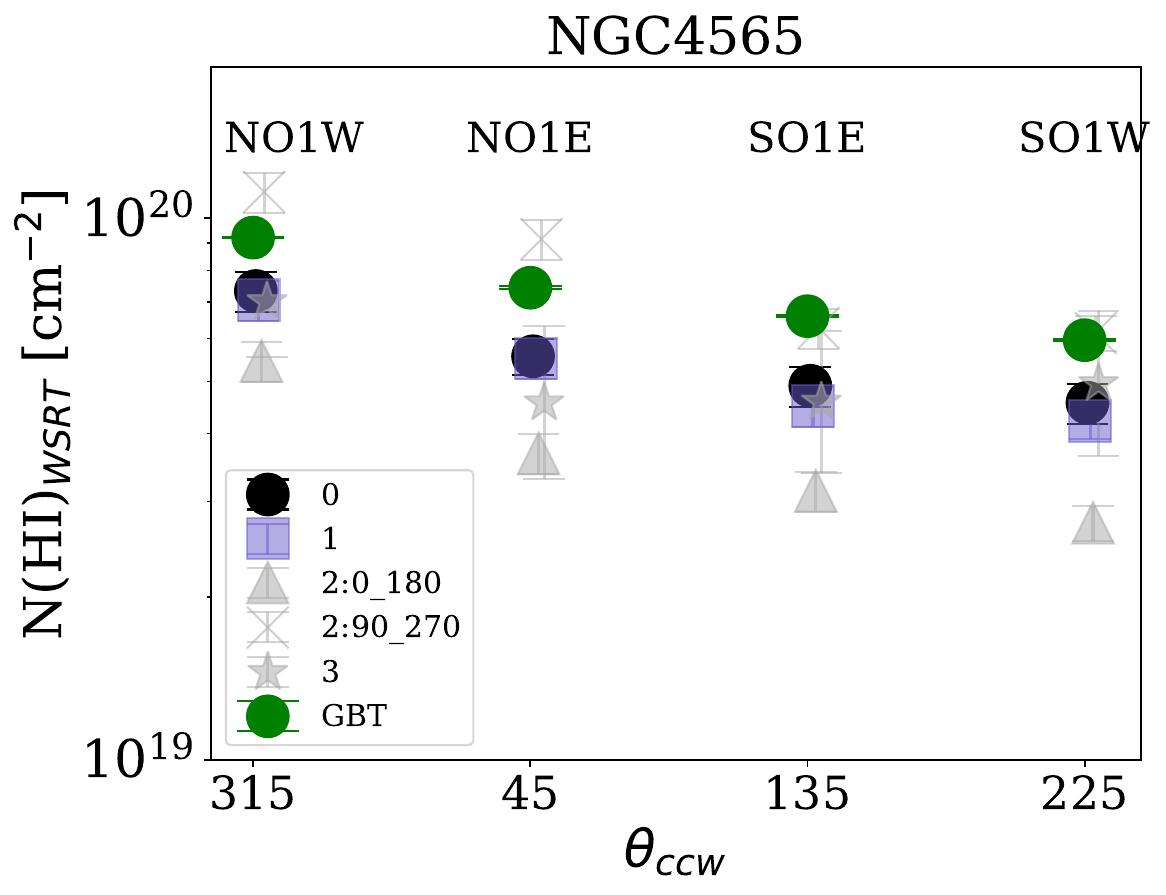}
    \caption{N(\hin) in WSRT convolved with the GBT beam using different methods described in detail in Appendix\,\ref{appdx:crossinstrument}. The corresponding N(\hin) values from the GBT for each pointing are marked with green circles, added for reference.} 
    \label{fig:appdx}
\end{figure*}

To calculate the N(\hin)$\rm_{WSRT}$ in Table \ref{tb:deriv}, referred to as ``Method 0" in Figure \ref{fig:appdx_allmeth}, we follow the same procedure as described in Section \ref{sec:comp_interfer}.
To confirm that the detected emission is truly diffuse circumgalactic gas rather than unaccounted disk contamination, we test different methods outlined in detail below. Method 1 tests whether we lose true emission in the masking process. Methods 2 and 3 constrain the systematic uncertainties resulting from the azimuthally asymmetric GBT beam. Comparison among these methods can be visualized in \ref{fig:appdx}.

\textbf{Method 1 (Unmasked):} The HALOGAS data from WSRT is $\sim$3 orders of magnitude shallower than our GBT data. Because of this sensitivity difference, it is possible that the excess emission detected by GBT is not truly diffuse, but rather, small clouds hidden under the noise and removed in the masking process. To test this, we convolve the unmasked HALOGAS cubes with the circularized GBT beam and follow the same steps as discussed earlier in this section and shown as the second panel from the top in Figure \ref{fig:appdx_allmeth}. The results from the unmasked cubes are plotted in Figure \ref{fig:appdx_allmeth} as the blue square points labeled ``unmasked". We find the unmasked results are consistent with results from ``Method 0". We feel confident that we are not missing any true emissions from HALOGAS in the masking process.

\textbf{Method 2 (`0 180/ 90 270'; GBT Beam Rotated position angles):} Because the off-axis positions demonstrate 2-3 times higher N(\hin) than the principal axes, it is important to investigate the possible effects of GBT beam orientation. The GBT beam has azimuthally asymmetric sidelobes (see Figure 1 from \citetalias{Pingel2018}). If, during a given observation, the beam were to be in one unique orientation for the entirety of an observation (e.g., oriented such that, because of the beam and sidelobe geometry, it was maximally sensitive along the minor axis), then we would expect to be most sensitive along that specific orientation. Although we had sufficiently long observations (3-4 hours), it is possible that we did not pass through all possible orientations of the beam, which could cause the circularized beam to under- or overestimate axes, which in turn incorrectly estimate the N(\hin) from the GBT beam-convolved WSRT column density.  To test the effect of the GBT beam asymmetry, we rotate the GBT beam along the position angle of the galaxy, position angle + 90$^\circ$, position angle + 180$^\circ$, and position angle + 270$^\circ$, to replicate the most extreme beam orientations shown in the second panel from the bottom of Figure \ref{fig:appdx_allmeth}. The position angles are 22$^\circ$ and 136$^\circ$ for NGC\,891 and NGC\,4565, respectively. 

We also note that the NGC\,4565 CGM pointings required an additional 7\,kms$^{-1}$ $\rm {v_{offset}}$ on top of the central 7\,kms$^{-1}$ $\rm {v_{offset}}$, meaning a total velocity adjustment of 14\,kms$^{-1}$ for NO1E and NO1W CGM pointings only. Before applying the scaling techniques (Appendix \ref{appdx:velocorrect}), we average the WSRT spectra obtained by convolving with the GBT beam rotated at position angle +90$^\circ$, position angle + 270$^\circ$, and position angle 0$^\circ$ and 180$^\circ$. This provides us with two spectra that correspond to the main lobe of the GBT beam aligned with the major axis (position angle, position angle +180$^\circ$) and the minor axis (position angle + 90$^\circ$, position angle +270$^\circ$). We apply the same process as described in \S\ref{sec:datared}, following the velocity adjustments (Appendix \ref{appdx:velocorrect}), T$_{\mathrm{B}}$ scaling, and sigma clipping. Depending on the orientation of the rotation, we would overestimate (or underestimate) any contamination from the major axes as the maximal sensitivity of the main beam aligns with the major axes. Since the off-axis pointings are $\sim$45 degrees between the principal axes pointings, it is essentially random whether or not the 0$^\circ$ and 180$^\circ$ rotation would maximize (or minimize), or the 90$^\circ$ and 270$^\circ$ would maximize (or minimize). As a consequence of this, we would overestimate the N(\hin) at some pointings and underestimate the N(\hin) in others, which we can see is true in Figure \ref{fig:appdx} where orange squares correspond to the position angle and position angle +180$^\circ$ averaged, and red squares correspond to position angle + 90$^\circ$, and position angle + 270$^\circ$ averaged. More importantly, the average of the 0$^\circ$, 180$^\circ$, 90$^\circ$, and 270$^\circ$ provides a conservative central estimate which aligns within 1$\sigma$ of the main result (``Method 0"). As a result, we are confident that the beam asymmetries are sufficiently accounted for in our results. 

\textbf{Method 3 (MUL; Median, upper, lower beam):}  The circularized GBT beam model and the uncertainties associated with it (see Table \ref{tb:deriv}) account for the varying sensitivity and change in antenna position. Instead of averaging the GBT beam response, we convolve the masked HALOGAS cube with three GBT beams: the median (i.e., 50th percentile), 84th percentile, and 16th percentile beam, see bottom panel of Figure \ref{fig:appdx_allmeth}. The median GBT beam serves as the base value. The upper and lower systematic uncertainty resulting from the GBT beam asymmetry is found by N(\hin)$_{84\%}$ - N(\hin)$_{50\%}$ and N(\hin)$_{50\%}$ - N(\hin)$_{16\%}$, respectively. The statistical uncertainty remains the same as before, the systematic uncertainty becomes the quadrature sum of 1) uncertainty in masking from ``Method 0" and 2) the upper (N(\hin)$_{84\%}$ - N(\hin)$_{50\%}$) and lower (N(\hin)$_{50\%}$ - N(\hin)$_{16\%}$) uncertainties.  We find the results are within 1$\sigma$ of the original results (Figure \ref{fig:appdx}), where the values from this attempt are labeled ``MUL" as stars in light gray. 

\section{Comparison with previous observations}
\label{appdx:p18}

The GBT data cubes of NGC\,891 and NGC\,4565 in \citetalias{Pingel2018} achieve a 5$\sigma$ sensitivity of $2\times 10^{18}$ \cmsq over a 20 kms$^{-1}$ linewidth in their spatial maps. In \cite{Das2020b}, we compared our deep stare GBT observations along minor axis pointings with those cubes and found mostly consistent results. However, the N(\hin) toward off-axis pointings is systematically $\approx$2--3 times higher than the N(\hin) toward principal axis pointings in both our GBT data and the HALOGAS data (see Figure\,2).  Here, we test if our measurements along the off-axis pointings are consistent with the GBT cubes in \citetalias{Pingel2018}. 

\begin{figure*}
\centering
\includegraphics[width=0.45\linewidth]{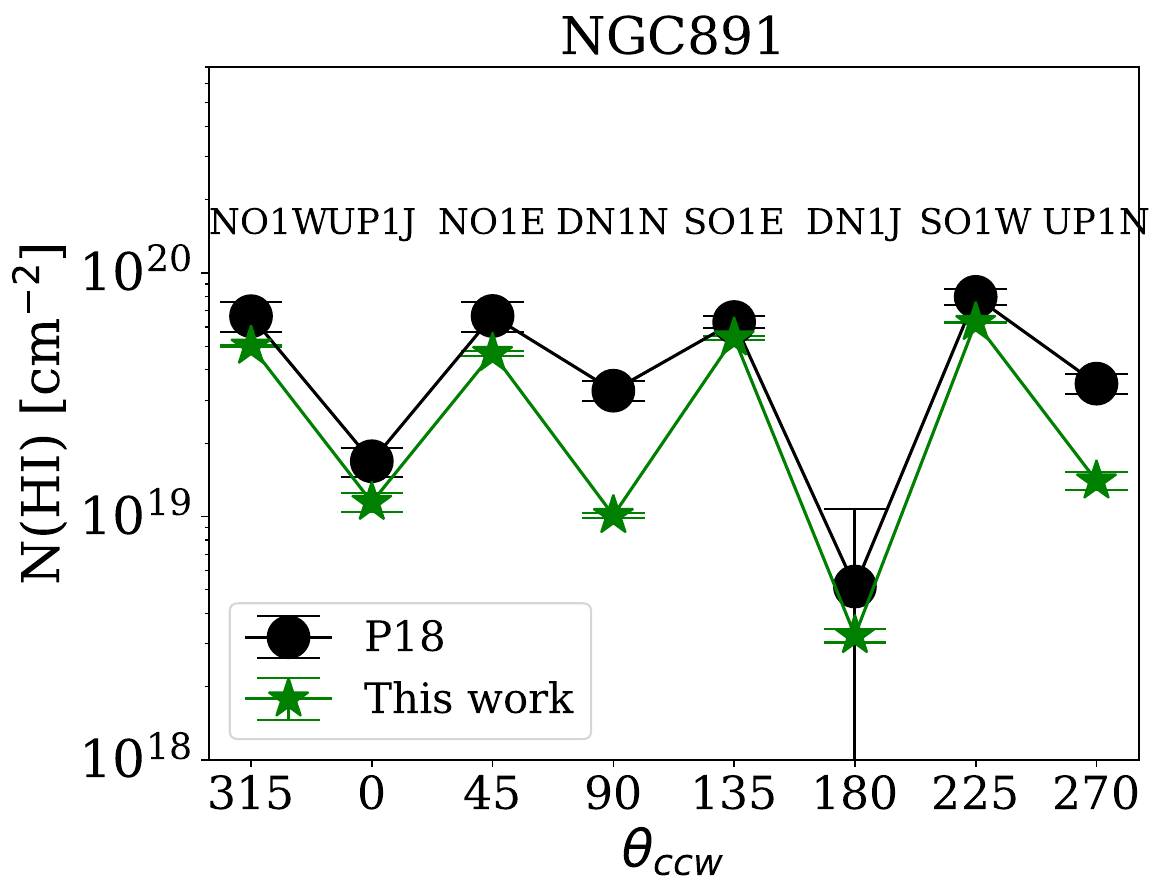}   \includegraphics[width=0.45\linewidth]{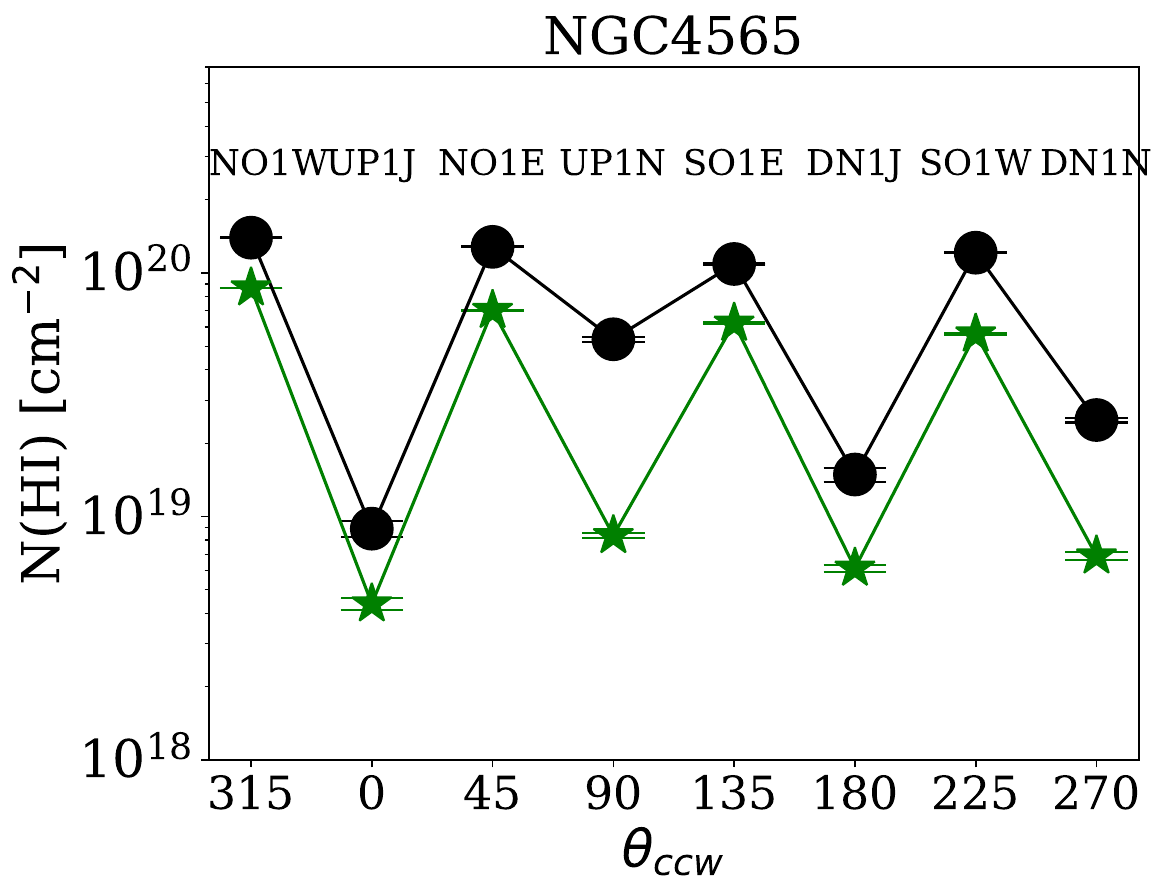}
\caption{Comparison of column densities from this work (green stars) with those extracted from GBT maps in \citetalias{Pingel2018} (black circles). Spectral extraction from \citetalias{Pingel2018} cubes is described in Appendix\,\ref{appdx:p18}.}
\label{fig:p18}
\end{figure*}

We extract the spectra from the cubes through the following steps. First, we determine the global noise value by extracting all values beyond r=30' and fitting the negative half of the histogram of the pixel values with a Gaussian described in further detail in  \citetalias{Das2024a} Appendix B. We then mask the \citetalias{Pingel2018} cube at the $5\sigma$ level using the global noise. We then create a subcube with an aperture size of 9.65 $'$ and compute its mean brightness temperature and velocity profiles. This approach differs from \citetalias{Pingel2018}, which did not apply an aperture constraint and instead summed over the entire field of view. The spectra were integrated over the same velocity range ($\Delta \mathrm{v_{emit}}$) as this work, then the \citetalias{Pingel2018} pointings were scaled by a single factor of NHI$_{P18,cent}$ /NHI$_{us,cent}$. The uncertainties in the \citetalias{Pingel2018} column density are computed as the quadrature sum of the uncertainty from varying the masking threshold (4–5$\sigma$ and 5–6$\sigma$) and the root-mean-square (RMS) noise of the signal-free part of the spectrum of the baseline (see D24 Appendix A). We plot the column densities from \citetalias{Pingel2018} spectra in Figure \ref{fig:p18}.

\citetalias{Pingel2018} similarly detects 2--3 times larger N(\hin) in the off-axis positions compared to the principal axes. This excess would reasonably be missed in the azimuthally averaged results originally presented in \citetalias{Pingel2018}.  Also, \citetalias{Pingel2018} values are systematically greater than ours, likely due to a larger effective aperture size in the mapping data (9.65$'$) compared to our deep stare (9.1$'$). This larger effective beam size in mapping is a consequence of gridding the individual GBT observations into the final cube using a convolution function of Gaussian-tapered circular Bessel functions, discussed in detail in \cite{Mangum2007}. We conclude the excess along the off-axis is \textit{true} and not an uncharacterized systematic effect.

\section{Verifying cross-pointing Contamination}
\label{appdx:beam_contam}

\begin{figure*}
\centering
\includegraphics[width=0.45\linewidth]{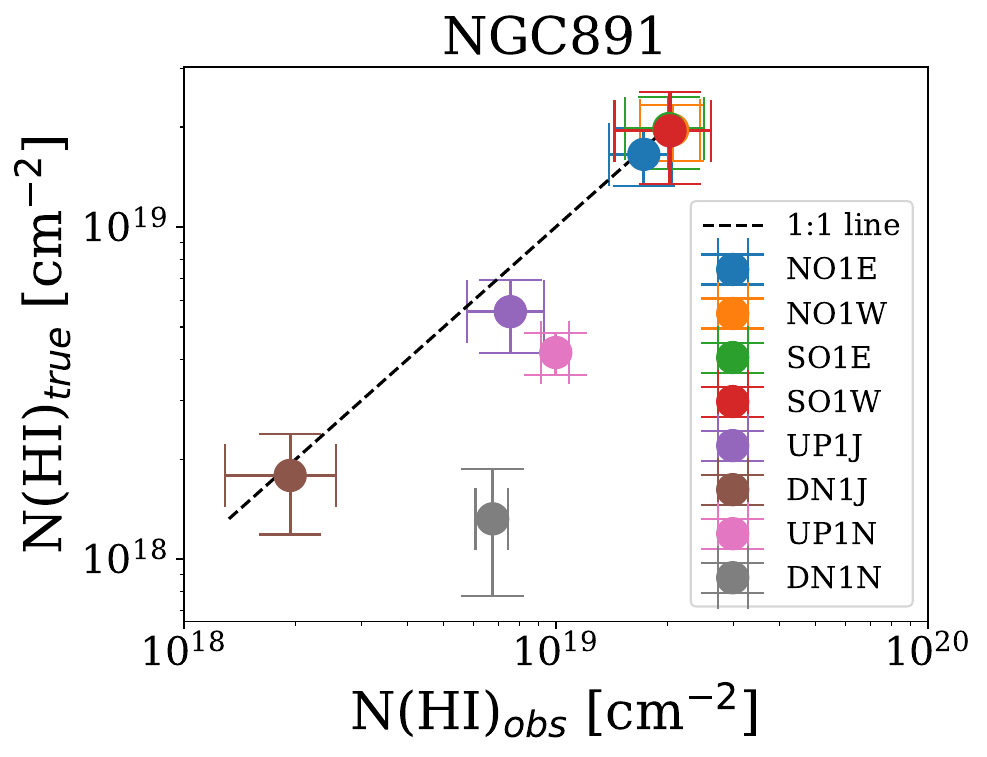}
\includegraphics[width=0.45\linewidth]{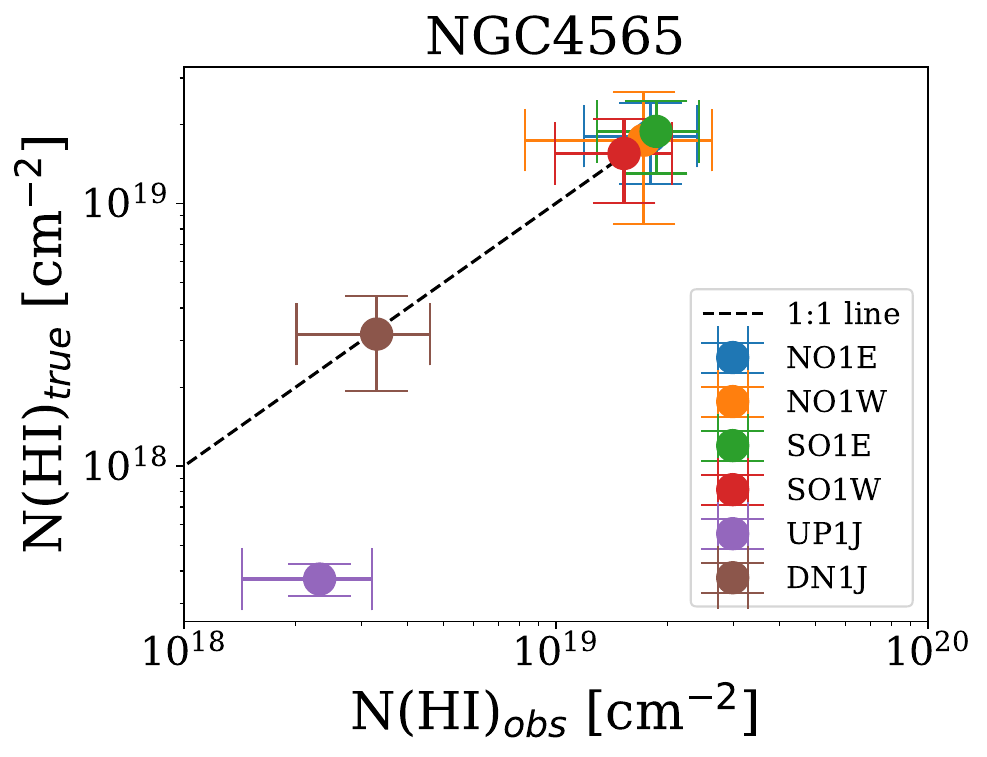}
\caption{N(HI)$_{\mathrm{obs}}$ is the observed N(HI)$_{\mathrm{CGM}}$ (Table \ref{tb:deriv}). N(HI)$_{\mathrm{true}}$ is the effectively deconvolved ``true" emission, discussed in Appendix \ref{appdx:beam_contam}. The black dashed line indicates the case of N(HI)$_{\mathrm{true}}$ =N(HI)$_{\mathrm{obs}}$.}
    \label{fig:contam}
\end{figure*}

All of our GBT pointings are at least one GBT FWHM away from each other and the edge of the \hi disk (see \S2.2). However, due to the sidelobes, the GBT beam extends far beyond the nominal FWHM of $\approx$9.1$'$. As a result, contamination from nearby pointings may be spuriously counted as the true emission at a given pointing. To quantify and correct for this, we use the following approach. First, we determine the angular separations from all possible pointing (N) pairs, resulting in an N$\times$N matrix. To calculate the contamination matrix (also N$\times$N), we extract the value of the circularized and normalized GBT beam at the pixel coordinate corresponding to the previously calculated angular separation between pointing pairs. 
\begin{equation}\label{eq:contam}
C_{\mathrm{true}} \;=\; F^{-1} \, C_{\mathrm{obs}} \,
\end{equation}
We use equation \ref{eq:contam} to solve for the ``true" emission, assuming all emissions originate from \hi clouds that are sufficiently small to be effectively approximated as a point source located at the sky position of each pointing. 
The input for the C$_{\mathrm{obs}}$ array is the N(\hin)$\rm_{CGM}$ values at each pointing (see Table\,2 and Figure\,3). To calculate the uncertainties, we consider C$_{\mathrm{obs,upper}}$ = C$_{\mathrm{obs}}$ + C$_{\mathrm{obs,err}}$, C$_{\mathrm{obs,lower}}$= C$_{\mathrm{obs}}$ - C$_{\mathrm{obs,err}}$. We then solve for the C$_{\mathrm{true,upper}}$ and C$_{\mathrm{true,lower}}$ as previously with equation \ref{eq:contam}. The uncertainties in C$_{\mathrm{true}}$ are calculated as C$_{\mathrm{true,upper-err}}$ = C$_{\mathrm{true,upper}}$ -C$_{\mathrm{true}}$, C$_{\mathrm{true,lower-err}}$ = C$_{\mathrm{true}}$ - C$_{\mathrm{true,lower}}$. 

We compare C$_{obs}$ and C$_{true}$ in Figure\,G1. The ``true" values are nearly consistent with (within $2\sigma$ of) the one-to-one line (C$_{obs}$ = C$_{true}$) or yield unphysical results (negative values from UP\,1N and DN\,1N of NGC\,4565, not plotted). The unphysical negative values indirectly tell us that the point source approximation is invalid for these pointings. In reality, N(\hin)$\rm_{CGM}$ consists of multiple low N(\hin) clouds with different sizes at different locations, as well as diffuse beam-filling extended \hin. While we have undertaken the first step to deconvolve single-dish deep stare data, solving for the \textit{true} circumgalactic \hi distribution from GBT data alone would require an involved process considering all physical possibilities of \hi morphology, size, and location; this is beyond the scope of this paper. 

Thus, the main takeaway from this investigation is that the systematic excess in N(\hin) toward off-axis pointings is not caused by the sidelobe contamination from surrounding CGM pointings, and that the distribution of circumgalactic \hi is azimuthally asymmetric.

\bibliography{sample631}{}
\bibliographystyle{aasjournal}

\end{document}